\documentstyle[12pt]{article}

\let\Large=\normalsize

\newcommand{\be}{\begin{equation}}
\newcommand{\ee}{\end{equation}}
\newcommand{\ba}{\begin{array}{c}}
\newcommand{\ea}{\end{array}}

\begin{document}

\begin{titlepage}
\begin{flushright}
CERN TH 6521/92\\
CPT-92 / P.2710\\
\end{flushright}
\vspace{2cm}
\begin{center}

\begin{Large}
\bf
NAMBU JONA-LASINIO LIKE MODELS\\
AND\\
THE LOW ENERGY EFFECTIVE ACTION OF QCD
\\
\end{Large}

\vspace*{1cm}
        {\bf Johan Bijnens}\\
         CERN, CH-1211 Geneva 23, Switzerland\\[1cm]
         {\bf Christophe Bruno} and {\bf Eduardo de Rafael}\\
         Centre de Physique Theorique\\
         CNRS - Luminy, Case 907\\
         F 13288 Marseille Cedex 9, France
\end{center}
\begin{abstract}
We present a derivation of the low energy effective action of an
extended Nambu Jona-Lasinio (ENJL)
model to $O(p^4)$ in the chiral counting.
Two alternative scenarios are considered on how the ENJL model could
originate as a low energy approximation to QCD.
The low energy effective Lagrangian we derive includes the usual
pseudoscalar Goldstone modes, as well as the lower scalar, vector and
axial-vector degrees of freedom. By taking appropriate limits, we recover
most of the effective low-energy models discussed in the literature; in
particular the gauged Yang-Mills vector Lagrangian, the Georgi-Manohar
constituent quark-meson model, and the QCD effective action approach
model. Another property of the ensuing effective Lagrangian is that it
incorporates most of the short-distance relations which follow from QCD.
(We derive these relations in the presence of all possible gluonic
interactions to leading order in the $1/N_c$-expansion.)
Finally the numerical predictions are compared to the experimental values
of the low energy parameters
\end{abstract}
\vskip 1cm
\begin{flushleft}
CERN-TH 6521/92\\
CPT-92 / P.2710\\
May   1992
\end{flushleft}
\end{titlepage}

\section{INTRODUCTION.}

\quad The original model of Nambu and Jona-Lasinio \cite{1} is
based on an analogy with
 superconductivity,
and it was proposed as a dynamical model of the strong
interactions between
nucleons and pions. In this model the pions appear as the
massless composite bosons associated with the dynamical
spontaneous breakdown of  the chiral
symmetry of the initial Lagrangian. This pioneering work
has had an enormous impact in the
development of modern particle physics and the standard model
in particular.

The idea that in QCD with three light flavours $u$, $d$ and $s$,
a mechanism ``somewhat like'' the one originally proposed by Nambu and
Jona-Lasinio may be at the origin of the dynamics responsible for the
spontaneous chiral symmetry breaking of the flavour group
$SU(3)_L \times SU(3)_R$ to $SU(3)_V$ is an appealing one. Several authors have
investigated this assumption with various claims of success in reproducing the
 features of low
energy hadron phenomenology. (Refs. \cite{3a} to \cite{3g}.)
{}From a theoretical point of view, however,
very little is known at present on
how an effective four-fermion interaction
\`a la Nambu Jona-Lasinio could be triggered within the framework of QCD.
Two alternative mechanisms
can be envisaged. One may first consider the formal
integration over gluon
fields in the path integral formulation of the generating
functional for the
Green's functions of quark-currents, and then truncate the
resulting non-local quark effective action to its lowest dimensional
interactions by invoking some ``local approximation.''
Alternatively,
one may consider a renormalization \`a la Wilson
where the quark and gluon degrees of freedom
have been integrated out down to a scale $\Lambda_{\chi}$
of the order of the spontaneous chiral symmetry breaking scale. The new
effective QCD-Lagrangian will indeed have local four-fermion couplings
normalized to the $\Lambda_{\chi}$-cut-off scale.
The assumption then is that
higher dimension operators play ``no fundamental role''
in the description of
the low energy physics. Within this alternative, one is still left with a
functional integral over the low frequency modes of the quark and gluon fields.

The QCD effective action approach proposed in Ref. \cite{13} and applied
successfully to non-leptonic $\Delta S = 1$ decays in Ref. \cite{14},
as well as
to the calculation of the imaginary part (terms of $O(p^6)$) of the QCD
effective action \cite{15},
and to the pion electromagnetic mass-difference
\cite{16},
falls in the second alternative described above. As we shall later
discuss (see also Ref. \cite{17}) the constituent chiral quark mass term

\be\label{1}
-M_Q (\bar {q}_L U^{\dagger} q_R + \bar {q}_R U q_L),
\ee

\noindent which in the approach of Ref. \cite{13} is added to the usual
QCD-Lagrangian, is equivalent to the mean-field
approximation of a Nambu Jona-Lasinio mechanism triggered by a four quark
operator of the type $\sum_{a,b} (\bar q_R^a q_L^b)(\bar q_L^b q_R^a)$,
where $a,b$ denote $SU(3)$ flavour indices and
summation over colour degrees
of freedom within each bracket is understood. Here

\be\label{2}
q_L \equiv {1\over 2} (1-\gamma_5) q(x)
\,\,\,\hbox{and}\,\,\,
q_R \equiv {1\over 2} (1+\gamma_5) q(x),
\ee

\noindent with $\bar q$ the flavour triplet of Dirac spinors ($\bar u (x) =
u^{\dagger}(x) \gamma^0$).

\be\label{3}
\bar{q} \equiv (\bar{u}(x),\bar{d}(x),\bar{s}(x)).
\ee

\noindent The $3\times 3$
matrix $U$ in (\ref{1}) is the unitary matrix which  collects
the Goldstone modes, i.e., the pseudoscalar degrees
of freedom ($\pi$, $K$ and
$\eta$) of the hadronic spectrum. Under chiral $SU(3)_L \times SU(3)_R$
transformations $U \to g_R U g_L^{\dagger}$ and (\ref{1}) is therefore
invariant. The mass parameter $M_Q$ provides a regulator of the infra-red
behaviour of the low energy effective action when the quark fields are
integrated out in the presence of a gluonic field background. The term in
 (\ref{1})
was proposed in Ref. \cite{13} as a phenomenological parametrization of
spontaneous chiral symmetry breaking.

The purpose of this article is to report on
a systematic study of the low energy
effective action of the extended
Nambu Jona-Lasinio model (ENJL) which, at
intermediate energies below or of the order
of a cut-off scale $\Lambda_{\chi}$,
is expected to be a good effective realization of the standard QCD
Lagrangian ${\cal L}_{QCD}$. The
Lagrangian in question is

\be\label{4}
{\cal L}_{QCD} \to {\cal L}_{QCD}^{\Lambda_{\chi}} + {\cal L}^{S,P}_{NJL} +
{\cal L}^{V,A}_{NJL} +O\left({1\over \Lambda_{\chi}^{4}}\right ),
\ee
\noindent with
\be\label{5}
{\cal L}^{S,P}_{NJL} = {8\pi^2 G_S(\Lambda_{\chi}) \over N_c \Lambda_{\chi}^2}
\sum_{a,b}(\bar q_R^a q_L^b)(\bar q_L^b q_R^a)
\ee
\noindent and
\be\label{6}
{\cal L}^{V,A}_{NJL} = - {8\pi^2 G_V(\Lambda_{\chi}) \over N_c
\Lambda_{\chi}^2}
\sum_{a,b}\left[(\bar q_L^a \gamma^{\mu} q_L^b)(\bar q_L^b \gamma_{\mu} q_L^a)
+ (L \to R) \right].
\ee

\noindent The couplings $G_S$ and $G_V$ are dimensionless quantities. In
principle they are calculable functions of the
$\Lambda_{\chi}$-cut-off scale.
In practice, the calculation requires knowledge of the non-perturbative
behaviour of QCD, and we shall
take $G_S$ and $G_V$ as independent unknown
constants. In choosing the forms
(\ref{5}) and (\ref{6}) of the four-quark
 operators, we have
only kept those couplings which are
allowed by the symmetries of the original
QCD-Lagrangian, and which are leading in the $1/N_c$-expansion,
where $N_c$
denotes the number of colours. With one inverse power of $N_c$ pulled-out;
$G_S$ and $G_V$ are constants of $O(1)$ in the large $N_c$ limit.

It is instructive to figure out how a four-fermion interaction like the one
above arises already at short distances in perturbative QCD. This is
illustrated in Fig. 1. The regulator replacement

\be\label{6a}
{1\over Q^2} \to \int_0^{1/\Lambda^2} d \tau e^{- \tau Q^2}
\ee

\noindent in the gluon propagator between the two interaction vertices in
Fig. 1a
leads to a local effective four-quark interaction ($g_s$ is the colour
coupling constant)

\be\label{6b}
{1\over \Lambda^2} g_s^2 \sum_{a} (\bar q(x) \gamma^{\mu}
{\lambda^{(a)}\over 2} q(x))(\bar q(x) \gamma_{\mu}
{\lambda^{(a)}\over 2} q(x))
\ee

\noindent as illustrated in Fig. 1b, which,
using Fierz identities and
neglecting subleading terms in $1/N_c$ reproduces the
form of the interaction
terms in \ref{5} and \ref{6} with

\be \label{6c}
G_S = 4 G_V = {g_s^2 \over 4\pi^2} N_c.
\ee

\noindent This perturbative estimate of $G_S$ and $G_V$ is however only valid
at $\Lambda$ scales sensibly larger than the $\Lambda_{\chi}$-scale at which
spontaneous chiral symmetry breaking takes place. In other words a reliable
calculation of $G_S$ and $G_V$ at realistic scales of $O(\Lambda_{\chi})$
necessarily involves non-perturbative dynamics.

The $\Lambda_{\chi}$ index in ${\cal L}_{QCD}^{\Lambda_{\chi}}$
means that
only the low frequency modes of the quark and gluon fields are
to be considered.
Of course, at the level where these low frequency gluonic effects
are ignored,
we are practically led to the first alternative we described above,
where the assumption is that ``all the relevant''
gluonic effects for low energy physics
can be absorbed in the new couplings $G_S$ and $G_V$.
We shall discuss how the two alternatives compare when confronted to
phenomenological predictions.

The paper is organized as follows.
In section 2 we describe the general frame work of QCD Green's functions
at low energies. In section 3 the form of the low energy effective
Lagrangians is given and its parameters derived from experiment.
The next section contains the discussion of spontaneous chiral
symmetry breaking in the extended Nambu Jona-Lasinio model including
gluonic interactions. Sections 5 and 6 are the main part of this work.
There we derive, respectively, the low energy effective action in the
two alternatives described above. We pay special attention to the
existence of relations between the low energy constants in the effective
action that are independent of the input parameters of the ENJL model.
Some of these relations remain true even after the inclusion
of gluonic interactions.
In section 7 we describe how this approach encompasses most other
models in the literature as various limits of parameters. In section
8 we give our numerical results using the formulas we derived earlier
and section 9 contains our main conclusions. A number of
technical remarks regarding the derivation of the low energy effective
action are collected in the appendices.

\section{QCD GREEN'S FUNCTIONS AT LOW ENERGIES.}

\quad In QCD with three light flavours u,d and s, the generating functional
for the Green's functions of vector, axial-vector, scalar and pseudoscalar
quark currents is defined by the vacuum-to-vacuum transition amplitude

\be\label{7}
e^{i\Gamma(v,a,s,p)} = \left<0 \left\vert T exp\left( i\int d^4 x
{\cal L}_{QCD}(x) \right) \right\vert 0\right>,
\ee

\noindent with ${\cal L}_{QCD}(x)$ the QCD Lagrangian in the presence of
external vector $v_{\mu}(x)$, axial-vector $a_{\mu}(x)$, scalar $s(x)$
and pseudoscalar $p(x)$ field sources; i.e.,

\be\label{8}
{\cal L}_{QCD}(x) = {\cal L}^o_{QCD} + \bar
q\gamma^{\mu}(v_{\mu}+\gamma_5a_{\mu})q - \bar q(s-i\gamma_5p)q,
\ee

\noindent where $q$ denotes the flavour triplet of Dirac spinors in (\ref{3});

\be\label{9}
{\cal L}^0_{QCD} = - {1\over 4} \sum^{8}_{a=1} G^{(a)}_{\mu\nu}
G^{(a)\mu\nu} + i\bar q \gamma^{\mu}(\partial_{\mu} + i G_{\mu})q,
\ee

\noindent and

\be\label{10}
G_{\mu} \equiv g_s \sum^{N_c^2-1}_{a=1} {\lambda^{(a)}\over 2}
G^{(a)}_{\mu} (x)
\ee

\noindent is the gluon field matrix in the fundamental $SU(N_c=3)_{colour}$
representation, with $G^{(a)}_{\mu\nu}$ the gluon field strength tensor

\be\label{11}
G^{(a)}_{\mu\nu} = \partial_{\mu} G^{(a)}_{\nu} - \partial_{\nu}
G^{(a)}_{\mu} - g_s f_{abc} G^{(b)}_{\mu} G^{(c)}_{\nu},
\ee

\noindent and $g_s$ the colour coupling constant $(\alpha_s = g^2_s/4\pi)$.
The external field sources are hermitian $3\times 3$ matrices in flavour space
and are colour singlets. The matrix field $s(x)$ contains
in particular the
quark mass matrix ${\cal M}=diag(m_u,m_d,m_s)$, i.e.,
$s(x)={\cal M}+\cdots$. The
other external matrix fields are traceless.

The Lagrangian ${\cal L}_{QCD}(x)$ is invariant under local chiral
$SU(3)_L \times SU(3)_R$ transformations; i.e., with $g_L,g_R \, \epsilon \,
SU(3)_L \times SU(3)_R$, and $q_{L,R}(x)$ defined as in (\ref{2})

\be\label{12}
q_L \rightarrow g_L (x) q_L
\,\,\,\hbox{and}\,\,\,
q_R \rightarrow g_R (x) q_R,
\ee

\be\label{13}
l_{\mu} \equiv v_{\mu} - a_{\mu} \rightarrow g_L l_{\mu} g_L^{\dagger} + i g_L
\partial_{\mu} g_L^{\dagger},
\ee

\be\label{14}
r_{\mu} \equiv v_{\mu} + a_{\mu} \rightarrow g_R r_{\mu} g_R^{\dagger} + i g_R
\partial_{\mu} g_R^{\dagger},
\ee

\noindent and

\be\label{15}
s+ip \rightarrow g_R (s+ip) g_L^{\dagger}.
\ee

\noindent In
fact, the Lagrangian ${\cal L}^0_{QCD}(x)$ is formally invariant
under global $U(3)_L \times U(3)_R$ transformations.
However, because of the
$U(1)_A$ anomaly this symmetry is broken from $U(3)_L \times U(3)_R$ to
$SU(3)_L \times SU(3)_R \times U(1)_V$. Since we shall
restrict ourselves to
states with zero baryon number the $U(1)_V$ factor
plays no role. We shall also
work in the large $N_c$-limit and restrict ourselves to
physics at leading
order in the $1/N_c$-expansion. In this limit the $U(1)_A$-anomaly
effects are absent.

It is convenient to use a path integral representation for the generating
functional $\Gamma(v,a,s,p)$,

\begin{displaymath}
e^{i\Gamma(v,a,s,p)} =
{1\over Z} \int {\cal D}G_{\mu} {\cal D}\bar{q} {\cal
D}q \hbox{exp} \left (i \int d^4 x {\cal L}_{\hbox{QCD}}
(q,\bar{q},G;v,a,s,p) \right )
\end{displaymath}

\be\label{16}
= \int {\cal D}G_{\mu} \hbox{exp}\left (-i\int d^4 x {1\over 4}
G^{(a)}_{\mu\nu} G^{(a)\mu\nu}\right ) {\cal D}\bar{q} {\cal
D}q \hbox{exp} \left (i \int d^4 x
\bar{q} i Dq \right ),
\ee

\noindent where $D$ denotes the Dirac operator

\be\label{17}
D = \gamma^{\mu}(\partial_{\mu} + i G_{\mu}) - i\gamma^{\mu}(v_{\mu} +
\gamma_5 a_{\mu}) + i(s-i\gamma_5 p).
\ee

\noindent The normalization factor $Z$ is fixed so that $\Gamma(0,0,0,0) = 0$.
The generating functional $\Gamma(v,a,s,p)$ is not
invariant under local chiral transformations due to the existence of
anomalies in the fermionic determinant.The structure of
the anomalous piece in
$\Gamma$ is known however from the work of Bardeen \cite{29}
and Wess and Zumino \cite{30}.

In QCD, the $SU(3)_L \times SU(3)_R$ symmetry in flavour space is expected
to be spontaneously broken down to $SU(3)_V$. In the limit of large $N_c$, this
has been proven to be the case under rather reasonable assumptions \cite{31}.
Numerical simulations of lattice regularized QCD also support this assumption
\cite{32}. According to Goldstone's theorem, there appears then an octet of
massless pseudoscalar particles $(\pi,K,\eta)$. The fields of these particles
can be conveniently collected in a $3 \times 3$ unitary matrix $U(\Phi)$ with
det$U =1$. Under local chiral transformations

\be\label{18}
U(x) \to g_R U(x) g_L^{\dagger}.
\ee

\noindent Whenever necessary, a useful parametrization for $U(\Phi)$, which
we shall adopt, is

\be\label{19}
U(\Phi) = \hbox{exp} \left( -i \sqrt 2 {\Phi(x) \over f_0} \right)
\ee

\noindent where $f_0 \simeq f_{\pi} = 93.2\, MeV$ and ($\stackrel{\rightarrow}
{\lambda}$ are Gell-Mann's
$SU(3)$ matrices with $tr\lambda_a \lambda_b=2\delta_{ab}$)

\be\label{20}
\Phi(x) = {\stackrel{\rightarrow}{\lambda} \over \sqrt 2}.
\stackrel{\rightarrow}{\Phi}(x) =
\left( \matrix {{ \pi^0 \over \sqrt 2}+{\eta \over \sqrt 6}&\pi^+&K^+\cr
                         \pi^-&{-\pi^0 \over \sqrt 2}+{\eta \over \sqrt
 6}&K^0\cr
                         K^-&\overline{K}^0&-2{\eta \over \sqrt 6}\cr} \right).
\ee

\noindent The $0^-$ octet $\Phi(x)$ is the ground state of the QCD hadronic
 spectrum.
There is a mass-gap from the ground state to the first massive multiplets with
$1^-$, $1^+$ and $0^+$ quantum numbers. The basic idea of the effective chiral
Lagrangian approach is that, in order to describe physics of the strong
interactions at low energies, it may prove more convenient to replace QCD by
an effective field theory which directly involves the pseudoscalar $0^-$ octet
fields; and, perhaps, the fields of the first massive multiplets
$1^-$, $1^+$ and $0^+$ as well.

The chiral symmetry of the underlying QCD theory implies that $\Gamma(v,a,s,p)$
 in eq. (\ref{16})
admits a low energy representation

\begin{displaymath}
e^{i \Gamma(v,a,s,p)} = {1\over Z}\int
{\cal D}U {\cal D}S {\cal D}V_{\mu} {\cal D}A_{\mu}
e^{i\int d^4x {\cal L}^R_{eff}(U,S,V_{\mu},A_{\mu};v,a,s,p)}
\end{displaymath}

\be\label{21}
= {1\over Z}\int {\cal D}U e^{i\int d^4x {\cal L}_{eff} (U;v,a,s,p)}
\ee

\noindent where the fields $S(x)$, $V_{\mu}(x)$ and $A_{\mu}(x)$ are those
associated with
the lowest massive scalar, vector and axial-vector particle
states of the hadronic spectrum. Both ${\cal L}^R_{eff}$ and
${\cal L}_{eff}$ are local Lagrangians which contain in principle
an infinite number of terms.
The hope is that, for sufficiently small energies
as compared to the spontaneous
chiral symmetry breaking scale $\Lambda_{\chi}$,
the restriction of ${\cal L}^R_{eff}$ and/or ${\cal L}_{eff}$
to a few terms with the lowest
chiral dimension should provide a sufficiently
accurate description of the low
energy physics. The success of this approach
at the phenomenological level is
by now confirmed by many examples\footnote{For recent reviews see e.g.
Refs. \cite{46} to \cite{48}.}.
Our aim here is to derive the effective Lagrangians ${\cal L}^R_{eff}$ and
${\cal L}_{eff}$ which follow from the Nambu Jona-Lasinio cut-off version
of QCD described in the introduction.

\section{LOW ENERGY EFFECTIVE LAGRANGIANS.}

\quad The purpose of this section is to summarize
briefly what is known at present   about
the low energy mesonic Lagrangians ${\cal L}^R_{eff}$ and
${\cal L}_{eff}$ from the chiral invariance properties of
${\cal L}_{QCD}$ alone.

The terms in ${\cal L}_{eff}$ with the lowest chiral dimension, i.e.,
$O(p^2)$ are
\be\label{22}
{\cal L}_{eff}^{(2)} = {1\over 4} f^2_0 \Big \{
tr D_{\mu} U D^{\mu} U^{\dagger} + tr(\chi U^{\dagger} + U^{\dagger} \chi)\Big
 \}
\ee

\noindent where $D_{\mu}$ denotes the covariant derivative

\be\label{23}
D_{\mu} U =\partial_{\mu} U -i(v_{\mu} + a_{\mu})U
+iU(v_{\mu} - a_{\mu})
\ee

\noindent

\be\label{24}
\chi = 2B_0(s(x)+ip(x)).
\ee

\noindent The constants $f_0$ and $B_0$ are not fixed by chiral symmetry
requirements. The constant $f_0$ can be obtained from $\pi \to \mu \nu$ decay,
 and
it is the same which appears in the normalization of the pseudoscalar field
matrix $\Phi(x)$ in (\ref{20}), i.e.,

\be\label{25}
f_0 \simeq f_{\pi} = 93.2\, MeV.
\ee

\noindent The constant $B_0$ is related to the vacuum expectation value

\be\label{26}
<0\vert \bar {q} q \vert 0>_{\vert q=u,d,s} = -f_0^2 B_0(1+O({\cal M})).
\ee

The terms in ${\cal L}_{eff}$ of $O(p^4)$ are also known. They have
been classified by Gasser and Leutwyler \cite{33}:

\be\label{27}
\matrix{
{\cal L}_{eff}^{(4)}(x)&= L_1(tr D_{\mu} U^{\dagger} D^{\mu} U)^2 +
L_2 tr(
D_{\mu} U^{\dagger} D_{\nu} U)tr(D^{\mu} U^{\dagger} D^{\nu} U)\hfill\cr
\cr
\hphantom{{\cal L}^{(4)}(x)}&+ L_3 tr(D_{\mu} U^{\dagger} D^{\mu} U D_{\nu}
 U^{\dagger}
D^{\nu} U)+L_4 tr( D_{\mu} U^{\dagger} D^{\mu} U)tr (\chi^{\dagger} U  +
U^{\dagger}\chi)\hfill\cr \cr
 \hphantom{{\cal L}^{(4)}(x)}&+ L_5 tr[D_{\mu}
U^{\dagger} D^{\mu} U (\chi^{\dagger} U + U^{\dagger} \chi)]+ L_6
 [tr(\chi^{\dagger} U + U^{\dagger} \chi)]^2  \hfill\cr
\cr
\hphantom{{\cal L}^{(4)}(x)}&+ L_7 [tr(\chi U - U^{\dagger} \chi)]^2 + L_8
tr(\chi^{\dagger} U \chi^{\dagger} U + \chi U^{\dagger} \chi
U^{\dagger})\hfill\cr\cr
\hphantom{{\cal L}^{(4)}(x)}&-iL_9 tr(F^{\mu\nu}_R D_{\mu} U D_{\nu}
U^{\dagger}
 +
F^{\mu\nu}_L D_{\mu} U^{\dagger} D_{\nu} U) + L_{10} tr(U^{\dagger}
 F^{\mu\nu}_R U
F_{L\mu\nu})\hfill\cr \cr
\hphantom{{\cal L}^{(4)}(x)}&+ H_1 tr(F^{\mu\nu}_R F_{R\mu\nu} +
F^{\mu\nu}_L F_{L\mu\nu}) + H_2 tr(\chi^{\dagger}\chi),\hfill
\cr}
\ee

\noindent where $F_{L \mu \nu}$ and $F_{R \mu \nu}$ are the external
field-strength tensors

\be\label{28}
F_{L \mu \nu} = \partial_{\mu} l_{\nu} - \partial_{\nu} l_{\mu}
- i [l_{\mu},l_{\nu}]
\ee

\be\label{29}
F_{R \mu \nu} = \partial_{\mu} r_{\nu} - \partial_{\nu} r_{\mu}
- i [r_{\mu},r_{\nu}]
\ee

\noindent associated with
the external left ($l_{\mu}$) and right ($r_{\mu}$)
field sources

\be\label{30}
l_{\mu} = v_{\mu} - a_{\mu}{\,\,\,},\,\,\,r_{\mu} = v_{\mu} + a_{\mu}.
\ee

\noindent The constants
$L_i$ and $H_i$ are again not fixed by chiral symmetry
requirements.
The $L_i$'s were phenomenologically determined in Ref. \cite{33}.
Since then, $L_{1,2,3}$
have been fixed more accurately using data from $K_{l4}$
\cite{34}.
The phenomenological values of the $L_i$'s which will be relevant for a
comparison  with our calculations, at a renormalization scale
$\mu = M_{\rho} =770\, MeV$, are collected in the first column in
Table \ref{table2}.

By contrast to ${\cal L}_{eff}$, which only has pseudoscalar fields as
physical degrees of freedom, the Lagrangian ${\cal L}^R_{eff}$ involves
chiral couplings of fields of massive $1^-$,$1^+$ and $0^+$ states, to
the Goldstone fields. The general method to construct these couplings was
described long time ago in Ref. \cite{35}.
An explicit construction of the
couplings for $1^-$,$1^+$ and $0^+$
fields can be found in Ref. \cite{36}. As
discussed in Ref. \cite{37},
the choice of fields to describe chiral invariant
couplings involving spin-1 particles is not unique and, when the vector modes
are integrated out, leads to ambiguities in the context of chiral perturbation
theory to $O(p^4)$ and higher. As shown in \cite{37},
these ambiguities are, however,
removed when consistency with the short-distance behaviour of QCD is
incorporated. The effective Lagrangian which we shall choose here to describe
vector couplings corresponds to the so called model II in Ref. \cite{37}.

The wanted ingredient for a non-linear representation of the chiral
$SU(3)_L \times SU(3)_R \equiv G$ group when dealing with matter fields is the
compensating $SU(3)_V$ transformation $h(\Phi,g_{L,R})$ which appears under
the action of the chiral group $G$ on the coset representative $\xi(\Phi)$ of
the $G/SU(3)_V$ manifold, i.e.,

\be\label{31}
\xi(\Phi) \to g_R \xi(\Phi) h^{\dagger}(\Phi,g_{L,R})
= h(\Phi,g_{L,R}) \xi(\Phi) g_L^{\dagger}
\ee

\noindent where $\xi(\Phi) \xi(\Phi) = U$ in the chosen gauge. This defines
the $3 \times 3$ matrix representation of the induced $SU(3)_V$ transformation.
Denoting the various matter $SU(3)_V$-multiplets by $R$ (octet) and $R_1$
(singlets), the non-linear realization of $G$ is given by

\be\label{32}
R \to h(\Phi,g_{L,R}) R h^{\dagger}(\Phi,g_{L,R})
\ee

\be\label{33}
R_1 \to R_1,
\ee

\noindent with the usual matrix notation for the octet

\be\label{34}
R = {1\over \sqrt 2} \sum^8_{i=1} \lambda^{(i)} R^{(i)}.
\ee

\noindent The vector field matrix $V^{\mu}(x)$ representing the $SU(3)_V$-octet
 of $1^-$ particles;
the axial-vector field matrix $A^{\mu}(x)$ representing $SU(3)_V$-octet of
$1^+$
 particles;
and the scalar field matrix $S(x)$  representing $SU(3)_V$-octet of $0^+$
 particles
are chosen to transform like $R$ in eq. (\ref{32}),
i.e.,($h \equiv h(\Phi,g_{L,R})$)

\be\label{35}
V_{\mu} \to h V_{\mu} h^{\dagger}\,\,;\,\,
A_{\mu} \to h A_{\mu} h^{\dagger}\,\,;\,\,
S \to h S h^{\dagger}.
\ee

The procedure now to construct the lowest order chiral Lagrangian
${\cal L}^R_{eff}$ is to write down all possible invariant couplings to first
non-trivial order in the chiral expansion which are linear in the $R$-fields
and to add of course the corresponding invariant kinetic couplings. It is
convenient for this purpose to set first the list of possible tensor structures
involving the $R$-fields which transform like $R$ in eq.(\ref{32}) under the
 action
of the chiral group $G$. Since the non-linear realization of $G$ on the octet
field $R$ is local, one is led to define a covariant derivative

\be\label{36}
d_{\mu}R = \partial_{\mu}R + [\Gamma_{\mu},R]
\ee

\noindent with a connection

\be\label{37}
\Gamma_{\mu} = {1 \over 2}\{\xi^{\dagger}[\partial_{\mu}-i(v_{\mu}+a_{\mu})]\xi
+ \xi[\partial_{\mu}-i(v_{\mu}-a_{\mu})]\xi^{\dagger}\}
\ee

\noindent ensuring the transformation property

\be\label{38}
d_{\mu}R \to h d_{\mu}R h^{\dagger}.
\ee

\noindent We can then define vector and axial-vector field strength tensors

\be\label{39}
V_{\mu\nu}=d_{\mu}V_{\nu}-d_{\nu}V_{\mu}\,\,\,{\hbox{and}}\,\,\,
A_{\mu\nu}=d_{\mu}A_{\nu}-d_{\nu}A_{\mu}
\ee

\noindent which also transform like $R$, i.e.,

\be\label{40}
V_{\mu\nu} \to h V_{\mu\nu} h^{\dagger}
\,\,\,{\hbox{and}}\,\,\,
A_{\mu\nu} \to h A_{\mu\nu} h^{\dagger}.
\ee

There is a complementary list of terms one can construct with the coset
 representative
$\xi(\Phi)$ and which transform homogeneously; i.e., like $R$ in (\ref{32}). If
 we restrict ourselves
to terms of $O(p^2)$ at most, here is the list:

\be\label{41}
\xi_{\mu} = i\{\xi^{\dagger}[\partial_{\mu}-i(v_{\mu}+a_{\mu})]\xi
- \xi[\partial_{\mu}-i(v_{\mu}-a_{\mu})]\xi^{\dagger}\}
= i\xi^{\dagger} D_{\mu} U \xi^{\dagger}=\xi_{\mu}^{\dagger},
\ee

\be\label{42}
\xi_{\mu}\xi_{\nu} \,\,\,{\hbox{and}}\,\,\, d_{\mu}\xi_{\nu},
\ee

\be\label{43}
\chi_{\pm} = \xi^{\dagger}\chi\xi^{\dagger} \pm \xi\chi^{\dagger}\xi,
\ee

\be\label{44}
f_{\mu\nu}^{\pm} = \xi F_{L \mu \nu}\xi^{\dagger} \pm \xi^{\dagger}F_{R \mu
 \nu}\xi.
\ee

\noindent Notice that $\Gamma_{\mu}$ in (\ref{37})
does not transform homogeneously, but
rather like an $SU(3)_V$ Yang-Mills field, i.e.,

\be\label{45}
\Gamma_{\mu} \to h \Gamma_{\mu} h^{\dagger} + h \partial_{\mu} h^{\dagger} .
\ee

The most general Lagrangian ${\cal L}^R_{eff}$ to lowest non-trivial order in
the chiral expansion is then obtained by adding to ${\cal L}^{(2)}_{eff}$ in
 eq.(\ref{22}) the
scalar Lagrangian

\be\label{46}
{\cal L}^S = {1 \over 2} tr\left(d_{\mu}S d^{\mu}S - M_S^2 S^2\right)
+ c_m tr\left(S \chi^+\right) + c_d tr\left(S \xi_{\mu} \xi^{\mu}\right);
\ee

\noindent the vector Lagrangian

\be\label{47}
{\cal L}^V = -{1 \over 4} tr\left(V_{\mu \nu} V^{\mu \nu} - 2M_V^2
V_{\mu}V^{\mu}\right) -{1 \over 2 \sqrt 2} \left[f_Vtr\left(
V_{\mu \nu} {f^{(+)}}^{\mu \nu}\right)
+  i g_Vtr\left(V_{\mu \nu}[\xi^{\mu},\xi^{\nu}]\right)\right]+
\cdots~,
\ee

\noindent and the axial-vector Lagrangian

\be\label{48}
{\cal L}^A = -{1 \over 4} tr\left(A_{\mu \nu} A^{\mu \nu} -
2M_A^2 A_{\mu}A^{\mu}\right)
-{1 \over 2 \sqrt 2} f_A tr\left(A_{\mu \nu} {f^{(-)}}^{\mu \nu}\right)+
\cdots ~,
\ee

\noindent The dots in ${\cal L}^V$ and ${\cal L}^A$ stand for other $O(p^3)$
 couplings which
involve the vector field $V^{\mu}$ and axial-vector field $A^{\mu}$ instead of
 the field-strength
tensors $V_{\mu\nu}$ and $A_{\mu\nu}$. They have been classified in Ref.
 \cite{37}. As discussed
there, they play no role in the determination of the $O(p^4)$ $L_i$ couplings
 when
the vector and axial-vector fields are integrated out.

The masses $M_V$, $M_S$ and $M_A$
and the coupling constants $c_m$, $c_d$,
 $f_V$, $g_V$ and $f_A$
are not fixed by chiral symmetry requirements. They can be determined
 phenomenologically as it
was done in Ref. \cite{36}.
Since later on we shall only calculate masses and
 couplings in the chiral limit, we
identify $M_V$, $M_S$ and $M_A$ to those of non-strange particles of the
corresponding multiplets, i.e.,

\be\label{49}
M_V=M_{\rho}=770\,MeV\,\,\, ; \,\,\, M_S=M_{a_0}=983\,MeV
\ee

\noindent and

\be\label{50}
M_A=M_{a_1}=1260 \pm 30\,MeV.
\ee

\noindent The couplings $f_V$ and $g_V$ can be then determined from the decay
 $\rho^0 \to e^+ e^-$
and $\rho \to \pi \pi$ respectively, with the result

\be\label{51}
\vert f_V \vert = 0.20\,\,\,{\hbox{and}}\,\,\,\vert g_V \vert = 0.090.
\ee

\noindent The decay $a_1 \to \pi \gamma$ fixes the coupling $f_A$ to

\be\label{52}
\vert f_A \vert = 0.097 \pm 0.022,
\ee

\noindent where the error is due to the experimental error in the determination
of the partial width \cite{49}, $\Gamma(a_1 \to \pi \gamma) = (640 \pm 246)
\,keV$. For the scalar couplings $c_m$ and $c_d$, the decay rate $a_0 \to
\eta \pi$ only fixes the linear combination \cite{36}

\be\label{53}
\vert c_d + {2m_{\pi}^2 \over M_{a_0}^2 - m_{\eta}^2 - m_{\pi}^2} c_m \vert =
 (34.3 \pm 3.3)\,MeV.
\ee

\noindent In confronting these results to theoretical predictions, one should
keep in mind that they have not been corrected for the effects of
chiral loop contributions.

\section{SPONTANEOUS CHIRAL SYMMETRY BREAKING A LA NAMBU JONA-LASINIO.}

\quad Following the standard procedure of introducing auxiliary fields, one
can rearrange the Nambu Jona-Lasinio cut-off version of the QCD Lagrangian in
an equivalent Lagrangian which is only quadratic in the quark fields. For this
purpose we shall introduce three complex $3\times 3$ auxiliary field matrices
$M(x)$, $L_{\mu}(x)$ and $R_{\mu}(x)$; the so called collective field
variables, which under the chiral group $G$ transform as

\be\label{54}
M \to g_R M g_L^{\dagger}
\ee

\be\label{55}
L_{\mu}\to g_L L_{\mu} g_L^{\dagger}\,\,\,{\hbox{and}}\,\,\,R_{\mu}\to g_R
 R_{\mu} g_R^{\dagger}.
\ee

\noindent We can then write the following identities:
\begin{displaymath}
 \hbox{exp} \,i\int d^4x {\cal L}_{NJL}^{S,P}(x) =
\end{displaymath}
\be\label{56}
\int {\cal D} M  \,  \hbox{exp} \,i\int d^4x  \left\{ -\left(\bar q_L
 M^{\dagger} q_R
+ \, h.c. \right) - {N_c \Lambda_{\chi}^2 \over 8\pi^2
 G_S(\Lambda_{\chi})}tr(M^{\dagger} M)\right\};
\ee

\noindent and

\begin{displaymath}
 \hbox{exp} \,i\int d^4x {\cal L}_{NJL}^{V,A}(x) =
\end{displaymath}
\be\label{57}
\int {\cal D}L_{\mu}{\cal D}R_{\mu}  \hbox{exp}  \,i\int d^4x
\left\{ \bar q_L \gamma^{\mu} L_{\mu} q_L + {N_c \Lambda_{\chi}^2 \over 8\pi^2
 G_V(\Lambda_{\chi})}
\left[{1 \over4} tr L^{\mu}L_{\mu} + (L \to R)\right\}\right],
\ee

\noindent where ${\cal L}_{NJL}^{S,P}(x)$ and ${\cal L}_{NJL}^{V,A}(x)$ are
the four-fermion  Lagrangians in (\ref{5}) and (\ref{6}).

By polar decomposition

\be\label{58}
M = U \tilde H = \xi H \xi,
\ee

\noindent with $U$ unitary and $\tilde H$ (and $H$) hermitian. From the
 transformation laws of $M$ and
$\xi$ in eqs. (\ref{54}) and (\ref{31}),
it follows that $H$ transforms homogeneously, i.e.,

\be\label{59}
H \to h(\Phi,g_{L,R}) H h^{\dagger}(\Phi,g_{L,R}).
\ee

\noindent The path integral measure in eq. (\ref{56})
can then also be written as

\begin{displaymath}
\exp \,\,i\int d^4x {\cal L}_{NJL}^{S,P}(x) =
\end{displaymath}
\be\label{60}
\int {\cal D} \xi {\cal D} H \, \hbox{exp}\,\, i\int d^4x  \left\{
-\left(\bar q_L \xi^{\dagger} H \xi^{\dagger} q_R + \bar q_R \xi H \xi
 q_L\right)
- {N_c \Lambda_{\chi}^2 \over 8\pi^2 G_S(\Lambda_{\chi})}tr H^2\right\}.
\ee

\noindent We are interested in the effective action
 $\Gamma_{eff}(H,\xi,L_{\mu},R_{\mu};v,a,s,p)$
defined in terms of the new auxiliary fields $H$,$\xi$,$L_{\mu}$, $R_{\mu}$;
and in the presence of the external field sources $v_{\mu}$, $a_{\mu}$,$s$ and
$p$, i.e.,

\begin{displaymath}
e^{i \Gamma_{eff}(H,\xi,L_{\mu},R_{\mu};v,a,s,p)} =
{1 \over Z}
\int {\cal D}G_{\mu} \hbox{exp} \left (-i \int d^4 x{1\over 4}G^{(a)}_{\mu\nu}
G^{(a)\mu\nu}\right) \times
\end{displaymath}

\begin{displaymath}
\hbox{exp} \,i \int d^4 x \left\{
{N_c\Lambda_{\chi}^2 \over 8\pi^2 G_V(\Lambda_{\chi})}{1
 \over4}[tr(L^{\mu}L_{\mu})+
tr(R^{\mu}R_{\mu})]
- {N_c \Lambda_{\chi}^2 \over 8\pi^2 G_S(\Lambda_{\chi})}
tr H^2\right\} \times
\end{displaymath}

\be\label{61}
\int {\cal D}\bar{q} {\cal D}q
\hbox{exp}\,\, i \int d^4 x \left\{
\bar{q}{\cal D}_{QCD} q +
\bar q_L \gamma^{\mu}L_{\mu} q_L +
\bar q_R \gamma^{\mu}R_{\mu} q_R
-\left(\bar q_L \xi^{\dagger} H \xi^{\dagger} q_R + \bar q_R \xi H \xi
 q_L\right) \right\},
\ee

\noindent with ${\cal D}_{QCD}$ the QCD Dirac operator

\be\label{62}
{\cal D}_{QCD} = \gamma^{\mu} (\partial_{\mu} + i G_{\mu})
- i \gamma^{\mu}(v_{\mu}+\gamma_5a_{\mu}) + i(s-i\gamma_5p).
\ee

\noindent The integrand is now quadratic in the fermion fields, and the path
 integral over the quark
fields is the determinant of
the full Dirac-operator in the Lagrangian (see
 appendix for technical
details).

Here, we are looking for
translational invariant solutions which minimize the
 effective action, i.e.,

\be\label{63}
{\delta \Gamma_{eff}(H,...) \over \delta
H}\vert_{L_{\mu}=R_{\mu}=0,\xi=1,H=<H>;
v_{\mu}=a_{\mu}=s=p=0} = 0
\ee

\noindent where $<H> = \hbox{diag} (M_u,M_d,M_s)$. The minimum is reached when
 all the eigenvalues
of $<H>$ are equal, i.e.,

\be\label{64}
<H> = M_Q 1
\ee

\noindent and the minimum condition leads to the so called gap equation

\be\label{65}
Tr(x\vert D_E^{-1}\vert x)\vert_
 {L_{\mu}=R_{\mu}=0,\xi=1,H=M_Q;v_{\mu}=a_{\mu}=s=p=0}
= - 4 M_Q {N_c \Lambda_{\chi}^2 \over 16\pi^2 G_S(\Lambda_{\chi})}\int d^4 x,
\ee

\noindent where $D_E^{-1}$ denotes the full Dirac-operator in
euclidean space-time. The trace in the
l.h.s. gives the formal evaluation of the vacuum expectation value of the
 quark-bilinear

\be\label{66}
<\bar{\psi}\psi> \equiv <\bar{u}u> = <\bar{d}d> = <\bar{s}s>.
\ee

\noindent In the large $N_c$-limit,
and with neglect of the gluonic couplings in
${\cal L}^{\Lambda_{\chi}}_{QCD}$,
this trace can be evaluated in the cut-off
 theory, with
the result (see appendix)

\be\label{67}
<\bar {\psi} \psi> = -{N_c \over  16 \pi^2} 4\!M_Q^3 \Gamma(-1,{M_Q^2\over
 \Lambda_{\chi}^2})
\ee

\noindent where $\Gamma(-1,x)$ denotes the incomplete gamma function

\be\label{68}
\Gamma(n-2,x={M_Q^2 / \Lambda_{\chi}^2}) =
\int_{M_Q^2 / \Lambda_{\chi}^2}^{\infty}{dz \over z}e^{-z} z^{n-2} ;
 \,\,\,\,\, n=1,2,3,...\ .
\ee

\noindent The corresponding gap equation results in the constraint

\be\label{69}
{M_Q \over G_S(\Lambda_{\chi})} =  M_Q\left\{\exp({-{M_Q^2
 \over\Lambda_{\chi}^2}})  -
{M_Q^2 \over \Lambda_{\chi}^2}
\Gamma (0,{M_Q^2\over \Lambda_{\chi}^2})\right\}.
\ee

\noindent This is the same solution as the one from
the Schwinger-Dyson equation which in a
diagrammatic notation is written in Fig. 2.
In terms of conventional Feynman
 diagrams, the set
of diagrams which are summed in the leading large-$N_c$ approximation are
chains
 of fermion bubbles
as indicated in Fig. 3a;
as well as trees of chains, like in Fig. 3b; but not
 loops of chains, as in Fig. 4.
Loops of chains are next to leading in the $1/N_c$-expansion.

Equations (\ref{67}) and (\ref{69}) show the existence of two-phases with
 regards to chiral symmetry.
The unbroken phase corresponds to

\be\label{70}
M_Q = 0 \Rightarrow <\bar {\psi} \psi> = 0
\ee

\noindent The broken phase corresponds
to the possibility that as we decrease  the ultraviolet
cut-off $\Lambda$ down to $\Lambda_{\chi}$
the coupling $G_S(\Lambda)$ increases
 allowing for
solutions to eq. (\ref{69}) with $M_Q>0$ and
therefore $<\bar {\psi} \psi> \neq 0$ and  negative. In
this phase, the hermitian auxiliary field $H(x)$
develops a nonvanishing vacuum  expectation value,
which is at the origin of a constituent chiral quark mass term

\be\label{71}
-M_Q (\bar {q}_L U^{\dagger} q_R + \bar {q}_R U q_L)
\ee

\noindent in the effective Lagrangian. This is precisely the term which in the
approach of Ref. \cite{13}
was incorporated  as a phenomenological parametrization of
 spontaneous chiral symmetry breaking.

Gluonic interactions due to fluctuations below the cut-off scale
 $\Lambda_{\chi}$
can be incorporated phenomenologically as proposed in Ref.\cite{13};
i.e., keeping the contributions
from the vacuum expectation values of gluon fields
which are leading  in the $1/N_c$-expansion.
This leads to correction terms in eqs. (\ref{67}) and (\ref{69})
with the result \cite{38}:

\begin{displaymath}
<\bar {\psi} \psi> = -{N_c \over  16 \pi^2} 4M_Q^3 \Gamma(-1,{M_Q^2\over
 \Lambda_{\chi}^2})
\end{displaymath}

\be\label{72}
-{1 \over 12}{< {\alpha_S \over \pi} GG> \over M_Q} \Gamma(1,{M_Q^2\over
 \Lambda_{\chi}^2})
-{1 \over 360}{\alpha_S \over \pi}g_S{<GGG> \over M_Q^3}\Gamma(2,{M_Q^2\over
 \Lambda_{\chi}^2})+...,
\ee

\noindent where the gluon condensates are understood as averages over frequency
 modes below the
cut-off scale $\Lambda_{\chi}$.

These gluonic corrections can be reabsorbed by an appropriate change of
the $\Lambda_{\chi}$-scale and a redefinition of the coupling constant $G_S$.
For this particular case, the two alternative mechanisms described in the
introduction are therefore equivalent.

\section{THE LOW ENERGY EFFECTIVE ACTION OF THE NAMBU JONA-LASINIO CUT-OFF
VERSION OF QCD.}

\subsection{The Mean Field Approximation.}

\quad We shall first discuss a particular case of
 $\Gamma_{eff}(H,\xi,L_{\mu},R_{\mu};v,a,s,p)$ as
defined in eq. (\ref{61}). It is the case corresponding to the mean field
 approximation, where

\be\label{73}
H(x) = <H> = M_Q {\LARGE\bf 1},
\ee

\noindent and where we set

\be\label{74}
L_{\mu} = R_{\mu} = 0.
\ee

\noindent The effective action
$\Gamma_{eff}(M_Q,\xi,0,0;v,a,s,p)$ coincides
 then with the one calculated in Ref. \cite{13},
except that the regularization  of the UV-behaviour
 is different.
In Ref. \cite{13}, the regularization which is used is the $\zeta$-function
 regularization. In a
cut-off theory, like the one we have now, we have a physical UV-cut-off
 $\Lambda_{\chi}$; and
the regularization must explicitly exhibit this
$\Lambda_{\chi}$-dependence. In  the calculations
reported here we have used a proper-time regularization (see appendix for
details).
The results, to a first approximation where low frequency gluonic
 terms are ignored, are
as follows:

\be\label{75}
f_0^2 = {N_c\over 16\pi^2} 4\,M_Q^2 \Gamma(0,{M_Q^2\over \Lambda_{\chi}^2})
\ee

\noindent and

\be\label{76}
f_0^2 B_0 = - <\bar {\psi} \psi> =
{N_c \over  16 \pi^2} 4\,M_Q^3 \Gamma(-1,{M_Q^2\over \Lambda_{\chi}^2})
\ee

\noindent for the lowest $O(p^2)$-couplings of the low energy effective
 Lagrangian in (\ref{22}).

For the $O(p^4)$-couplings which exist in the chiral limit we find

\be\label{77}
L_2=2L_1={N_c\over 16\pi^2}{1\over 12}\Gamma(2,{M_Q^2\over \Lambda_{\chi}^2}),
\ee

\be\label{78}
L_3={N_c\over 16\pi^2}{1\over 6}\left[\Gamma(1,{M_Q^2\over \Lambda_{\chi}^2})
- 2\Gamma(2,{M_Q^2\over \Lambda_{\chi}^2})\right]
\ee

\noindent for the four derivative terms; and

\be\label{79}
L_9={N_c\over 16\pi^2}{1\over 3}\Gamma(1,{M_Q^2\over \Lambda_{\chi}^2}),
\ee

\be\label{80}
L_{10}=-{N_c\over 16\pi^2}{1\over 6}\Gamma(1,{M_Q^2\over \Lambda_{\chi}^2})
\ee

\noindent for the two couplings involving external fields. If one lets
${M_Q^2\over \Lambda_{\chi}^2} \to 0$, then $\Gamma(n,0)=\Gamma(n)=(n-1)$! for
 $n\geq 1$, and these
results coincide with those previously obtained in Refs. \cite{8a}, \cite{8b},
\cite{10b}, \cite{13} and \cite{50} to \cite{50c}.

When terms proportional to the quark mass matrix ${\cal M}$ are kept, there
 appear four new
$L_i$-couplings (see eq. (\ref{27})). With

\be\label{81}
\rho = {M_Q \over \vert  B_0 \vert} = {M_Q f_0^2 \over \vert <\bar{\psi} \psi>
 \vert},
\ee

\noindent the results we find for these new couplings are

\be\label{82}
L_4=0
\ee

\be\label{83}
L_5={N_c\over 16\pi^2} {\rho \over 2}
\left[\Gamma(0,{M_Q^2\over \Lambda_{\chi}^2})-\Gamma(1,{M_Q^2\over
 \Lambda_{\chi}^2})\right]
\ee

\be\label{84}
L_6=0
\ee

\be\label{85}
L_7={N_c \over 16\pi^2}{1 \over 12}\left[-\rho \Gamma(0,{M_Q^2\over
 \Lambda_{\chi}^2})
+{1 \over 6}\Gamma(1,{M_Q^2\over \Lambda_{\chi}^2})\right]
\ee

\be\label{86}
L_8=-{N_c \over 16\pi^2} {1 \over 24}\left[6\rho(\rho - 1)\Gamma(0,{M_Q^2\over
 \Lambda_{\chi}^2})
+\Gamma(1,{M_Q^2\over \Lambda_{\chi}^2}) \right].
\ee

\noindent If we identify $\Gamma(0,{M_Q^2\over \Lambda_{\chi}^2})
\equiv \hbox{log}{\mu^2 \over M_Q^2}$, and take the limit
$\Gamma(n\geq 1,{M_Q^2\over \Lambda_{\chi}^2} \to 0)$ these results coincide
 then with those obtained in
Ref. \cite{13}.
(Notice that $\rho$ is twice the parameter $x$ of Ref. \cite{13}.)

The fact that $L_4=L_6=0$ and $L_2=2L_1$, is more general than the model
 calculations we are
discussing. As first noticed by Gasser and Leutwyler \cite{33}, these are
 properties of the large
$N_c$ limit. The contribution we find for $L_7$ is in fact non-leading in the
 $1/N_c$-expansion. The
result above is
entirely due to the use of the lowest order equations of motion
 (see the erratum to Ref. \cite{13}).
In the presence of the $U_A(1)$-anomaly, $L_7$ picks up a
 contribution from the
$\eta^{\prime}$-pole and becomes $O(N_c^2)$, \cite{33}.

Finally, we shall also give the results for the $H_1$ and $H_2$ coupling
 constants of terms which
only involve external fields:

\be\label{87}
H_1=-{N_c \over16\pi^2}{1 \over 12}\left[2\Gamma(0,{M_Q^2\over
 \Lambda_{\chi}^2})
-\Gamma(1,{M_Q^2\over \Lambda_{\chi}^2})\right]
\ee

\be\label{88}
H_2={N_c \over16\pi^2}{1 \over 12}\left[6\rho^2\Gamma(-1,{M_Q^2\over
 \Lambda_{\chi}^2})
-6\rho(\rho + 1)\Gamma(0,{M_Q^2\over \Lambda_{\chi}^2}) +
\Gamma(1,{M_Q^2\over \Lambda_{\chi}^2})\right]
\ee

\subsection{Beyond the Mean Field Approximation.}

\quad In full generality,

\be\label{89}
H(x) = M_Q {\LARGE\bf 1} + \sigma(x);
\ee

\noindent and the effective action
$\Gamma_{eff}(H,\xi,L_{\mu},R_{\mu};v,a,s,p)$
has a non-trivial dependence on the auxiliary field variables $\sigma(x)$,
$L_{\mu}(x)$ and $R_{\mu}(x)$. It is convenient to trade the auxiliary left and
 right
vector field variables $L_{\mu}$ and $R_{\mu}$ which were introduced in eq.
 (\ref{57}), by the new
vector fields

\be\label{90}
W_{\mu}^{\pm} = \xi L_{\mu} \xi^{\dagger} \pm \xi^{\dagger} R_{\mu} \xi.
\ee

\noindent From the transformation properties in eqs. (\ref{31}) and (\ref{55})
 it follows that
$W_{\mu}^{\pm}$ transform homogeneously; i.e.,

\be\label{91}
W_{\mu}^{\pm} \to h(\Phi,g) W_{\mu}^{\pm} h^{\dagger}(\Phi,g).
\ee

\noindent We also find it convenient to rewrite the effective action in eq.
 (\ref{61}) in a basis of
constituent chiral quark fields

\be\label{92}
Q = Q_L + Q_R\,\,\,\hbox{and}\,\,\,\bar Q = \bar Q_L + \bar Q_R
\ee

\noindent where

\be\label{93}
Q_L = \xi q_L\,,\,\bar Q_L = \bar q_L \xi^{\dagger}\,;
\,Q_R = \xi^{\dagger} q_R \,,\,\bar Q_R = \bar q_R \xi,
\ee

\noindent which under the chiral group $G$, transform like

\be\label{94}
Q \to h(\Phi,g) Q \,\,\,\hbox{and}\,\,\,\ \bar Q \to \bar Q
h(\Phi,g)^{\dagger}.
\ee

\noindent In this basis, the linear terms (in the auxiliary field variables) in
 the r.h.s. of eq.
(\ref{61}) become

\be\label{95}
 \bar q_L \gamma^{\mu}L_{\mu} q_L + \bar q_R \gamma^{\mu}R_{\mu} q_R
-\left(\bar q_L \xi^{\dagger} H \xi^{\dagger} q_R + \bar q_R \xi H \xi
 q_L\right)
\ee

\be\label{96}
\to \bar Q \left( - H + {1\over 2} \gamma^{\mu}W_{\mu}^+ -
{1\over 2} \gamma^{\mu}\gamma_5W_{\mu}^- \right) Q.
\ee

\noindent The effective action
 $\Gamma_{eff}(M_Q,\xi,\sigma,W_{\mu}^{\pm};v,a,s,p)$ in terms of the
new auxiliary field variables, and in euclidean space is then

\begin{displaymath}
e^{\Gamma_{eff}(M_Q,\xi,\sigma,W_{\mu}^{\pm};v,a,s,p)} =
\end{displaymath}
\begin{displaymath}
\hbox{exp}\left( - \int d^4 x \left\{
{N_c \Lambda_{\chi}^2 \over 8\pi^2 G_S(\Lambda_{\chi})}tr H^2 +
{N_c \Lambda_{\chi}^2 \over 16\pi^2 G_V(\Lambda_{\chi})}{1 \over 4}tr(W_{\mu}^+
 W_{\mu}^+ +
W_{\mu}^- W_{\mu}^-)\right\}\right)\times
\end{displaymath}
\be\label{98}
{1 \over Z}
\int {\cal D}G_{\mu} \hbox{exp} \left (- \int d^4 x{1\over
 4}G^{(a)}_{\mu\nu}G^{(a)}_{\mu\nu}\right)
\int {\cal D}\bar{Q} {\cal D}Q \hbox{exp} \int d^4 x \bar{Q}{\cal D}_E Q
\ee

\noindent where ${\cal D}_E$ denotes the euclidean Dirac operator

\be\label{99}
{\cal D}_E = \gamma_{\mu} \nabla_{\mu} - {1 \over 2}(\Sigma - \gamma_5
\Delta) - H(x)
\ee

\noindent with $\nabla_{\mu}$, the covariant derivative

\be\label{100}
\nabla_{\mu}=\partial_{\mu} + iG_{\mu} + \Gamma_{\mu} - {i \over 2} \gamma_5
(\xi_{\mu} - W_{\mu}^{(-)}) - {i \over 2} W_{\mu}^{(+)}
\ee

\noindent and

\be\label{101a}
\Sigma = \xi^{\dagger} \cal{M} \xi^{\dagger} + \xi \cal{M}^{\dagger} \xi
\ee
\be\label{101b}
\Delta = \xi^{\dagger} \cal{M} \xi^{\dagger} - \xi \cal{M}^{\dagger}
 \xi.
\ee

\noindent The quantities $G_{\mu}$, $\Gamma_{\mu}$ and $\xi_{\mu}$ are those
 defined in eqs.
(\ref{10}), (\ref{37}) and (\ref{41}).

At this stage, it is worth pointing  out a formal symmetry which is useful to
 check explicit
calculations. We can redefine the external vector-field sources via

\be\label{102}
l_{\mu} \to l_{\mu}^{\prime}=l_{\mu} + L_{\mu}
\ee

\be\label{103}
r_{\mu} \to r_{\mu}^{\prime}=r_{\mu} + R_{\mu}
\ee

\noindent and

\be\label{104}
{\cal M} \to {\cal M}^{\prime}(x) = {\cal M} + \xi \sigma(x) \xi.
\ee

\noindent The Dirac operator ${\cal D}_E$ in eq. (\ref{99}),
when reexpressed  in terms of the
``primed'' external fields reads

\be\label{105}
{\cal D}_E = \gamma_{\mu} (\partial_{\mu} + {\cal A}_{\mu}) + M
\ee

\noindent with

\be\label{106}
{\cal A}_{\mu} = iG_{\mu} + \Gamma_{\mu}^{\prime} - {i \over 2} \gamma_5
 \xi_{\mu}^{\prime}
\,\,\,\hbox{and}\,\,\,\
M = - {1 \over 2}(\Sigma^{\prime} - \gamma_5 \Delta^{\prime}) - M_Q,
\ee

\noindent and where $\Gamma_{\mu}^{\prime}$, $\xi_{\mu}^{\prime}$,
$\Sigma^{\prime}$ and $\Delta^{\prime}$ are the
same as in eqs. (\ref{37}),
 (\ref{41}),(\ref{101a}) and (\ref{101b})   with $l_{\mu} \to
l_{\mu}^{\prime}$,
$r_{\mu} \to r_{\mu}^{\prime}$
 and
${\cal M} \to {\cal M}^{\prime}$. Formally, this is the
same Dirac operator as
 the one corresponding
to the ``mean   field approximation'' which we discussed
in the previous paragraph.
 In practice, it
means that once we have evaluated the formal effective action

\be\label{107}
\hbox{exp } \Gamma_{eff}({\cal A}_{\mu},M) =
\int {\cal D}\bar{Q} {\cal D}Q \hbox{exp} \int d^4 x \bar{Q}{\cal D}_E Q =
 \hbox{det} {\cal D}_E
\ee

\noindent we can easily get the new terms
involving the new auxiliary fields
 $L_{\mu}$, $R_{\mu}$
and $\sigma$ by doing the appropriate shifts. The formal evaluation of
$\Gamma_{eff}({\cal A}_{\mu},M)$ to $O(p^4)$
in the chiral expansion has been
 made by several authors, Refs. \cite{50} to \cite{50c}.
We reproduce the results in the appendix.

\subsection{The constant $g_A$ and resonance masses.}

\quad When computing the effective action $\Gamma_{eff}({\cal A}_{\mu},M)$ in
 eq.
 (\ref{107}) there appears
a mixing term proportional to $tr \xi_{\mu} {W^{(-)}}^{\mu}$. More precisely,
 one finds a
quadratic form in $\xi_{\mu}$ and $W^{(-)}_{\mu}$ (in Minkowski space-time):

\be\label{108}
\Gamma = \alpha <{W^{(-)}}_{\mu}W^{(-){\mu}}> + \beta
<\xi_{\mu}W^{(-){\mu}}> + \gamma <\xi_{\mu}\xi^{\mu}>
\ee

\noindent with

\be\label{109}
\alpha = {N_c \over 16\pi^2}\left({1\over 4}{\Lambda_{\chi}^2 \over G_V} +
M_Q^2\Gamma(0,{M_Q^2 \over \Lambda_{\chi}^2}) \right)
\ee

\be\label{110}
\beta = -2{N_c \over 16\pi^2}\Gamma(0,{M_Q^2 \over \Lambda_{\chi}^2}) M_Q^2
\,\,\,\hbox{and}\,\,\, \gamma = -{1\over 2} \beta
\ee

\noindent The field redefinition

\be\label{111}
{W^{(-)}}_{\mu} \to {\hat W^{(-)}}_{\mu} + (1-g_A)\xi_{\mu}
\ee

\noindent with

\be\label{112}
g_A = 1+{\beta \over 2\alpha}
\ee

\noindent diagonalizes the quadratic form. There is a very interesting physical
 effect due to this
diagonalization, which is that it redefines the coupling of the constituent
 chiral quarks to the
pseudoscalars. Indeed, the covariant derivative in eq. (\ref{100}) becomes

\be\label{113}
\nabla_{\mu}=\partial_{\mu} + iG_{\mu} + \Gamma_{\mu} - {i
\over 2} \gamma_5  (g_A \xi_{\mu} - \hat W_{\mu}^{(-)}) - {i \over 2}
 W_{\mu}^{(+)}.
\ee

\noindent We shall later come back to $g_A$ and its possible identification
with
 the $g_A$-coupling constant of the constituent
chiral quark model of Manohar and Georgi \cite{39}.

In the calculation of $\Gamma_{eff}({\cal A}_{\mu},M)$ we also encounter
kinetic
 like terms
for the fields $\hat W^{(-)}_{\mu}$ and $W^{(+)}_{\mu}$ :

\be\label{114}
-{N_c \over 16\pi^2}{1 \over 3}\Gamma(0,{M_Q^2 \over \Lambda_{\chi}^2})
{1 \over 4}tr (\partial_{\mu}W^{(+)}_{\nu} - \partial_{\nu}W^{(+)}_{\mu})
(\partial^{\mu}W^{(+){\mu}} - \partial^{\nu}W^{(+){\mu}}),
\ee

\noindent and

\be\label{115}
-{N_c \over 16\pi^2}{1 \over 3}[\Gamma(0,{M_Q^2 \over \Lambda_{\chi}^2})
- \Gamma(1,{M_Q^2 \over \Lambda_{\chi}^2})]
{1 \over 4}tr (\partial_{\mu}\hat W^{(-)}_{\nu} - \partial_{\nu}\hat
 W^{(-)}_{\mu})
(\partial^{\mu}\hat W^{(-){\nu}} - \partial^{\nu}\hat W^{(-){\mu}}).
\ee

\noindent Comparison with the standard vector and axial-vector kinetic terms
in eqs. (\ref{47}) and (\ref{48}), requires
a scale redefinition of the fields $W^{(+)}_{\mu}$ and $\hat W^{(-)}_{\mu}$ to
 obtain the correct
kinetic couplings, i.e.,

\be\label{116}
V_{\mu} = \lambda_V W^{(+)}_{\mu}
\,\,\,\,\,\,\,\,
A_{\mu} = \lambda_A \hat W^{(-)}_{\mu},
\ee

\noindent with

\be\label{117}
\lambda_V^2 = {N_c \over 16\pi^2} {1 \over 3}
\Gamma(0,{M_Q^2 \over \Lambda_{\chi}^2})
\ee

\noindent and

\be\label{118}
\lambda_A^2 ={N_c \over 16\pi^2} {1 \over 3}
\left[\Gamma(0,{M_Q^2 \over \Lambda_{\chi}^2}))-\Gamma(1,{M_Q^2 \over
 \Lambda_{\chi}^2}))\right].
\ee

\noindent This scale redefinition gives rise to mass terms (in Minkowski
 space-time)

\be\label{119}
{1 \over 2} M_V^2 tr(V_{\mu}V^{\mu})+ {1 \over 2}M_A^2 tr(A_{\mu}A^{\mu})
\ee

\noindent with

\be\label{120}
M_V^2 = {2\alpha +\beta \over \lambda_V^2}
\,\,\,\,\,\,\hbox{and}\,\,\,\,\,\,
M_A^2 = {2\alpha \over \lambda_A^2}.
\ee

The same comparison between the calculated kinetic and mass terms in the scalar
 sector with the
standard scalar Lagrangian in eq. (\ref{46}), requires the scale redefinition

\be\label{121}
S(x) = \lambda_S \sigma(x),
\ee

\noindent with

\be\label{122}
\lambda_S^2 = {N_c \over 16\pi^2} {2 \over 3} \left[3\Gamma(0,{M_Q^2 \over
 \Lambda_{\chi}^2})
- 2\Gamma(1,{M_Q^2 \over \Lambda_{\chi}^2}) \right].
\ee

\noindent The scalar mass is then

\be\label{123}
M_S^2 = {N_c \over 16\pi^2} {8M_Q^2  \over \lambda_S^2} \Gamma(0,{M_Q^2 \over
 \Lambda_{\chi}^2}).
\ee

\subsection{The couplings of the ${\cal L}^R_{eff}$-Lagrangian.}

\quad The Lagrangian in question is the one we have written in section 3 in
eqs. (\ref{46}), (\ref{47}) and (\ref{48}), based on
chiral symmetry requirements alone. These requirements  did not fix,
however,
the masses and the interaction couplings with the pseudoscalar fields and
external fields.
The results for the masses which we now find in the extended
Nambu Jona-Lasinio model are given by
eqs. (\ref{120}) and (\ref{123}) in the previous subsection.
These are the results in the limit where
low frequency gluonic interactions in
${\cal L}^{\Lambda_{\chi}}_{QCD}$ in eq. (\ref{4}) are neglected, i.e.,
the results
 corresponding to the first
alternative scenario we discussed in the introduction. For the other coupling
 constants, and also
in the limit where low frequency gluonic interactions are neglected,
the results  are the following:

\be\label{124}
{1\over 4}f_{\pi}^2 = {N_c\over 16\pi^2} M_Q^2 g_A\Gamma(0,{M_Q^2 \over
 \Lambda_{\chi}^2}),
\ee

\noindent instead of the mean
field approximation result in eq. (\ref{75});\footnote{This implicitly
changes the value of $B_0$ via eq. \protect{(\ref{76})}.}

\be\label{125}
f_V = \sqrt 2 \lambda_V\,\,\,\,\, , \,\,\,\,\,f_A = \sqrt 2 g_A \lambda_A;
\ee
\be\label{126}
g_V = {N_c \over 16\pi^2} {1 \over \lambda_V} {\sqrt 2 \over 6}
\left[(1-g_A^2) \Gamma(0,{M_Q^2 \over \Lambda_{\chi}^2}) +
2 g_A^2 \Gamma(1,{M_Q^2 \over \Lambda_{\chi}^2}) \right]
\ee

\noindent for the vector and axial-vector coupling constants in (\ref{47}) and
 (\ref{48}); and

\be\label{127}
c_m = {N_c \over 16\pi^2} {M_Q \over \lambda_S} \rho
\left[\Gamma(-1,{M_Q^2 \over \Lambda_{\chi}^2}) - 2 \Gamma(0,{M_Q^2 \over
 \Lambda_{\chi}^2}) \right],
\ee

\be\label{128}
c_d = {N_c \over 16\pi^2} {M_Q \over \lambda_S} 2 g_A^2\left[\Gamma(0,{M_Q^2
 \over \Lambda_{\chi}^2})
-\Gamma(1,{M_Q^2 \over \Lambda_{\chi}^2})\right]
\ee

\noindent for the scalar coupling constants in (\ref{46}).

There are a series of interesting relations between these results, which we
 collect below:

\be\label{129}
M_V^2 = {3 \over 2} {\Lambda_{\chi}^2 \over G_V(\Lambda_{\chi}^2)}
{1 \over \Gamma(0,{M_Q^2 \over \Lambda_{\chi}^2})},
\ee

\be\label{130}
M_A^2\left\{1- {\Gamma(1,{M_Q^2 \over \Lambda_{\chi}^2}) \over
\Gamma(0,{M_Q^2 \over \Lambda_{\chi}^2})}\right\} = M_V^2 + 6 M_Q^2,
\ee

\be\label{131}
g_A = 1+{\beta \over 2\alpha} = {f_V^2 M_V^2 \over f_A^2 M_A^2}g_A^2,
\ee

\noindent with the two solutions

\be\label{132}
g_A = 0 \,\,\,\hbox{and}\,\,\, g_A = {f_A^2 M_A^2 \over f_V^2 M_V^2};
\ee

\noindent and

\be\label{133}
f_V^2 M_V^2 = f_A^2 M_A^2 + f_{\pi}^2.
\ee

\noindent The last relation is the $1^{st}$ Weinberg sum rule \cite{40}.
Using this sum rule and the second solution for $g_A$, we also have that

\be\label{134}
g_A = 1 - {f_{\pi}^2 \over f_V^2 M_V^2}.
\ee

\noindent We find therefore that $g_A<1$. As we shall discuss in section 6, the
 two relations
in eqs. (\ref{133}) and (\ref{134}) remain valid in the presence of gluonic
 interactions; i.e., the
gluonic corrections do modify the explicit form of the calculation we have made
 of $f_{\pi}$, $f_V$,
$M_V$ and $g_A$, but they do it in such a way that eqs. (\ref{133}) and
 (\ref{134}) remain unchanged.

\subsection{The coupling constants $L_i$'s, $H_1$ and $H_2$ beyond the mean
 field approximation.}

\quad These coupling constants are now modified because no longer we have
 $g_A=1$.
 With the short-hand
notation

\be\label{135}
x = {M_Q^2 \over \Lambda_{\chi}^2},
\ee

\noindent the analytic expressions we find from the quark-loop
integration are the following:

\be\label{136}
 L_2 = 2\,L_1 = {N_c \over 16\pi^2}{1\over 24}\left[(1-g_A^2)^2\Gamma(0,x)
+ 4g_A^2 (1-g_A^2) \Gamma(1,x)+ 2g_A^4 \Gamma(2,x)\right],
\ee
\be\label{137}
 \tilde L_3 = {N_c \over 16\pi^2}{1\over 24}\left[-3(1-g_A^2)^2\Gamma(0,x)
+ 4\left(g_A^4 - 3g_A^2 (1-g_A^2)\right)\Gamma(1,x) -
 8g_A^4\Gamma(2,x)\right],\ee
\be\label{138}
L_4 =0,\ee
\be\label{139}
\tilde L_5 = {N_c\over 16\pi^2} {\rho \over 2}
 g_A^2\left[\Gamma(0,x)-\Gamma(1,x)\right],\ee
\be\label{140}
 L_6 = 0,\ee
\be\label{141}
L_7 = O(N_c^2), \ee
\be\label{142}
\tilde L_8 = -{N_c \over 16\pi^2} {1\over 24}\left[ 6 \rho (\rho - g_A)
\Gamma(0,x) + g_A^2\Gamma(1,x) \right],\ee
\be\label{143}
L_9 = {N_c \over 16\pi^2}{1\over 6}\left[(1-g_A^2)\Gamma(0,x) +2
 g_A^2\Gamma(1,x)\right],\ee
\be\label{144}
L_{10} = -{N_c \over16\pi^2}{1\over6}\left[(1-g_A^2)\Gamma(0,x) +
 g_A^2\Gamma(1,x)\right],\ee
\be\label{145}
H_1 = -{N_c \over16\pi^2}{1\over12}\left[(1+g_A^2)\Gamma(0,x) -
 g_A^2\Gamma(1,x)\right],\ee
\be\label{146}
\tilde H_2 = {N_c \over16\pi^2}{1 \over 12}\left[6 \rho^2 \Gamma(-1,x) -
6\rho (\rho + g_A)\Gamma(0,x) +g_A^2\Gamma(1,x)\right].\ee

\noindent Three of the $L_i$-couplings ($i=$ 3, 5 and 8) as well as $H_2$
receive explicit contributions from the integration of scalar
fields. This is why we write $L_i = \tilde L_i + L_i^S$, $i=$ 3, 5, 8 ;
$H_2 = \tilde H_2 + H_2^S$ with $\tilde L_i$, $\tilde H_2$ the contribution
from the quark-loop and $L_i^S$, $H_2^S$ from the scalar field.
The results for $L_1$, $L_2$ and $\tilde L_3$ agree with those
of Ref. \cite{58} where these couplings were obtained by integrating out the
constituent quark fields in the model of Manohar and Georgi \cite{39}. At
the level where possible gluonic corrections are neglected, the two
calculations are formally equivalent.
There also exists a recent calculation of the $L_i$-couplings in the literature
within the framework of an extended Nambu Jona-Lasinio model as we are
discussing here. Our results for $L_4$ to $L_{10}$ agree with those of
Ref. \cite{5}.

We note that between these results for the $L_i$'s, $H_1$ and
the results for couplings and masses of the  vector and axial-vector
Lagrangians which we obtained before there are the following interesting
 relations:

\be\label{147}
L_9 = {1 \over 2} f_V g_V,
\ee

\be\label{148}
L_{10} = -{1 \over 4} (f_V^2 - f_A^2)\,\,\,\hbox{and}\,\,\,
2\,H_1 = -{1 \over 4} (f_V^2 + f_A^2).
\ee

\noindent As we shall see in the next section,
these relations, like those in eqs. (\ref{133}) and (\ref{134}), are also valid
 in the presence of
gluonic interactions. The alerted reader will recognize that these relations
are
 precisely the QCD
short-distance constraints which,
as discussed in Ref. \cite{37}, are required to
 remove the ambiguities
in the context of chiral perturbation theory to $O(p^4)$ when vector and
 axial-vector degrees of freedom are integrated out.
They are the relations which follow from demanding
consistency  between the low
energy effective action of vector and axial-vector
mesons and the QCD short-distance behaviour of two-point functions and
three-point functions. It is rather remarkable that the simple ENJL
model we have been discussing
incorporates these constraints automatically.

There is a further constraint which was
also invoked in Ref. \cite{37}. It has to
do with the asymptotic
behaviour of the elastic meson-meson scattering, which in
QCD is expected to  satisfy the Froissart bound \cite{41}.
If that is the case, the authors of Ref. \cite{37} concluded that,
 besides the constraints already discussed, one also must have

\be\label{149}
L_1 = {1 \over 8} g_V^2\,\,;\,\, L_2 = 2 L_1 \,\,;\,\, L_3 = - 6
 L_1. \ee

\noindent As already mentioned,
the second constraint is a property of QCD in the large $N_c$-limit.
The first and third constraint however
are highly non-trivial. We observe that,
to the extent that $O(N_c g_A^4)$ terms can be neglected,
these constraints
are then also satisfied in the ENJL model we are
considering. Indeed, it follows from eqs. (\ref{136}), (\ref{137}) and
 (\ref{126}) that

\be\label{150}
{\tilde L}_3+6L_1 = {N_c \over 16 \pi^2} g_A^4 {1 \over 12}
[2\Gamma(1,x)-\Gamma(2,x)],
\ee

\be\label{151}
8L_1-g_V^2={N_c \over 16 \pi^2} g_A^4 {1 \over 3} [\Gamma(2,x)-
2{\Gamma(1,x)^2 \over \Gamma(0,x)}].
\ee

When the massive scalar field is integrated out \cite{36},
there is a further
 contribution to the
constants $L_3$, $L_5$, $L_8$ and $H_2$ with the results:

\be\label{152}
L_3^S = {c_d^2 \over 2 M_S^2} = {N_c \over 16\pi^2}{1 \over 4} g_A^4 {1 \over
 \Gamma(0,x)}
[\Gamma(0,x)-\Gamma(1,x)]^2,
\ee
\be\label{153}
L_5^S = {c_m c_d \over M_S^2} = {N_c \over 16\pi^2}{1 \over 4} \rho g_A^2 {1
 \over \Gamma(0,x)}
[\Gamma(-1,x)-2\Gamma(0,x)][\Gamma(0,x)-\Gamma(1,x)],
\ee
\be\label{154}
L_8^S = {c_m^2 \over 2 M_S^2} = {N_c \over 16\pi^2}{1 \over 16} \rho^2 {1 \over
 \Gamma(0,x)}
[\Gamma(-1,x)-2\Gamma(0,x)]^2,
\ee
\be\label{155}
H_2^S = 2L_8^S.
\ee

\noindent Our result for $L_3^S$ disagrees with the one found in Ref.
\cite{5}. Also, contrary to what is found in Ref. \cite{5}, there is no
contribution from scalar exchange to $L_2$.

It is interesting to point out that $\tilde L_5$, $L_5^S$ and
$\tilde L_8$, $L_8^S$ each depend explicitly on the parameter $\rho$.
This dependence however, disappears in the sums

\be \label{155a}
L_5 = \tilde L_5 + L_5^S = {N_c \over 16 \pi^2} {1\over 4}g_A^3
\left[\Gamma(0,x)-\Gamma(1,x)\right];
\ee
\noindent and
\be\label{155b}
L_8 = \tilde L_8 + L_8^S = {1 \over 4} f_{\pi}^2 {g_A \over 16 M_Q^2} -
{N_c \over 16\pi^2} {1\over 24} g_A^4 \Gamma(1,x).
\ee
\section{THE LOW ENERGY EFFECTIVE ACTION IN THE PRESENCE OF GLUONIC
INTERACTIONS.}

\quad The purpose of this section is to explore more in detail the second
 alternative
which we described in the introduction wherewith the four quark operator terms
in eqs. (\ref{5}) and (\ref{6}) are viewed as the leading result of a first
step
renormalization \`a la Wilson, once the quark and gluon degrees of freedom
have been integrated out down to a scale $\Lambda_{\chi}$. Within this
alternative, one is still left with a fermionic determinant which has to be
evaluated  in the presence of gluonic interactions due to fluctuations below
the $\Lambda_{\chi}$-scale. The net effect of these long distance gluonic
interactions is to modify the various incomplete gamma functions
$\Gamma(n,x={M_Q^2 \over \Lambda_{\chi}^2})$
which modulate the calculation of
the fermionic determinant in the previous sections, into new (a priori
incalculable) constants. We examine first, how many independent
unknown constants can appear at most.
Then, following the approach developed in Ref. \cite{13},
we shall proceed to an approximate calculation of the new
constants to order $\alpha_S N_c$.

\subsection{Book-keeping of (a priori) unknown constants.}

\quad The calculation of the effective action in the previous sections, has
been
organized as a power series in proper time (see the appendix for details).
This is the origin of the integrals of the type

\begin{displaymath}
\int_{1 / \Lambda^2_{\chi}}^{\infty}{d \tau \over \tau}
{1 \over 16 \pi^2 \tau^2} \tau^n e^{-\tau M_Q^2} =
{1 \over 16 \pi^2} {1 \over (M_Q^2)^{n-2}}
\int_{M_Q^2 / \Lambda^2_{\chi}}^{\infty}{dz \over z}e^{-z} z^{n-2}
\end{displaymath}
\be\label{156}
= {1 \over 16 \pi^2} {1 \over (M_Q^2)^{n-2}}
\Gamma(n-2,x={M_Q^2 / \Lambda^2_{\chi}}); \,\,\,\,\, n=1,2,3,...\,\,\,.
\ee

\noindent In the presence of a gluonic background, each term in the effective
 action
which originates on a fixed power of the proper time expansion of the heat
kernel, becomes now modulated by an infinite series in powers of colour
singlet gauge invariant combinations of gluon field operators. Eventually,
we have to take the statistical gluonic average over each of these series.
In practice, each different average becomes an unknown constant. If we limit
ourselves to terms in the effective action to $O(p^4)$ at most, there can only
appear a finite number of these unknown constants. We can make their
book-keeping by tracing back all the possible different types of terms which
can
appear.

In order to proceed further with this book-keeping,
it is convenient to rewrite the
operator $E$ (see eq. (\ref{A10}) in the appendix)
in the following short-hand notation

\be\label{157}
E= {\cal \bf S} + \gamma_{\mu} {\cal \bf V}_{\mu} +
\sigma_{\mu \nu} {\cal \bf R}_{\mu \nu}.
\ee

\noindent Clearly,

\be\label{158}
{\cal \bf S} \equiv {1 \over 4} {\Sigma^{\prime}}^2 - M_Q^2 -
{1 \over 4} {\Delta^{\prime}}^2-
{1 \over 8} \gamma_5 [ \Sigma^{\prime},\Delta^{\prime}],
\ee

\be\label{159}
{\cal \bf V}_{\mu} \equiv {i \over 4}\gamma_5
\{\xi_{\mu}^{\prime},\Sigma^{\prime}
- \gamma_5 \Delta^{\prime}\} + {1 \over 2} d^{\prime}_{\mu}(\Sigma^{\prime}
- \gamma_5 \Delta^{\prime}),
\ee

\be\label{160}
{\cal \bf R}_{\mu \nu} \equiv -{i \over 2} R_{\mu \nu}^{\prime}.
\ee

\noindent We now observe that terms with only one power of
${\cal \bf S}$, when calculated in the
limit $\alpha_S N_c \to 0$ can only appear modulated by the factor
$\Gamma(-1,x)$, since they necessarily come from $tr E$
which is in $H_1(x,x)$ (see eq. (\ref{A23})).
For these terms, the net effect of the gluonic interactions will be
to renormalize the factor $\Gamma(-1,x)$ into a new constant

\be\label{161}
\Gamma(-1,x) \to \Gamma(-1,x)(1+\gamma_{-1}),
\ee

\noindent with $\gamma_{-1}$ an unknown functional of gluonic averages.
This explains the meaning of eq. (\ref{162}).  The meaning of the others,
eqs. (\ref{163}) to (\ref{166}), is similar. E.g., the terms in eq.
(\ref{163}) proportional to $1+\gamma_{01}$ all come from the
${\bf V}_\mu {\bf V}_\mu$ in the $E^2$ part of $H_2$.
In the limit $\alpha_S N_c \to 0$ they are modulated by
$\Gamma(0,x)$.
All these terms
will be modulated by the same new unknown factor when including gluonic
effects. These we have absorbed in the free coefficient $\gamma_{01}$.

The net effect of the gluonic interactions is
to modulate the various terms in
$\Gamma_{eff}^{(i)}$, $i = 1,2,3,4$, as given in appendix \ref{appD},
equations (\ref{A28}) to (\ref{A32}),
in the following way:

\be\label{162} \Gamma_{eff}^{(1)} = {N_c \over
 16\pi^2}\Gamma(-1,x)(1+\gamma_{-1})
tr\left[2({1\over 4}{\Sigma^{\prime}}^2 - M_Q^2)\right];\ee

\begin{displaymath}
\Gamma_{eff}^{(2)} = {N_c \over 16\pi^2} \Gamma(0,x)
tr\left[{1\over 2} (1+\gamma_{01})\{{\xi^{\prime}}^{\mu},\Sigma^{\prime}\}^2 +
{i\over 4}(1+\gamma_{01})\{\xi^{\prime}_{\mu},\Sigma^{\prime}\}
{d^{\prime}}^{\mu}\Delta^{\prime}\right.
\end{displaymath}
\begin{displaymath}
- (1+\gamma_{02})({1\over 4}{\Sigma^{\prime}}^2 - M_Q^2)^2
+ {1\over 4}(1+\gamma_{01})d^{\prime}_{\mu}\Sigma^{\prime}
{d^{\prime}}^{\mu}\Sigma^{\prime}
\end{displaymath}
\be\label{163}
\left.-{1\over 12}(1+\gamma_{03})\left({f^{(+)}}^{\prime}_{\mu\nu}
{f^{(+)}}^{{\prime}\mu \nu}
+{f^{(-)}}^{\prime}_{\mu \nu}{f^{(-)}}^{{\prime}\mu \nu}\right)\right];
\ee

\begin{displaymath}
 \Gamma_{eff}^{(3)} = {N_c \over 16\pi^2} \Gamma(1,x)
 tr\left[
-{1 \over 16}(1+\gamma_{11})({1\over 4}{\Sigma^{\prime}}^2-M_Q^2)
\{\xi^{\prime}_{\mu},\Sigma^{\prime }\}^2
- {i \over 2}(1+\gamma_{12}){f^{(+)}}^{\prime}_{\mu
 \nu}{\xi^{\prime}}^{\mu}{\xi^{\prime}}^{\nu}\right.
\end{displaymath}
\be\label{164}
\left.+ {1 \over 6} (1+\gamma_{13})\left((d^{\prime}_{\mu}\xi^{\prime}_{\nu})^2
+ {1\over
2}(\xi^{\prime}_{\mu}{\xi^{\prime}}^{\mu}\xi^{\prime}_{\nu}{\xi^{\prime}}^{\nu}
 +
\xi^{\prime}_{\mu}\xi^{\prime}_{\nu}{\xi^{\prime}}^{\mu}
{\xi^{\prime}}^{\nu})\right)
-{1\over 6}(1+\gamma_{14})d^{\prime}_{\mu}\Sigma^{\prime}
{d^{\prime}}^{\mu}\Sigma^{\prime}\right],
\ee

\noindent with

\begin{displaymath}
(d^{\prime}_{\mu}\xi^{\prime}_{\nu})^2 =
-{1\over
4}[\xi^{\prime}_{\mu},\xi^{\prime}_{\nu}][{\xi^{\prime}}^{\mu},{\xi^{\prime}}^{
 \nu}]
+ i {f^{(+)}}^{\prime}_{\mu \nu} {\xi^{\prime}}^{\mu} {\xi^{\prime}}^{\nu}
+ {1\over 2}{f^{(-)}}^{\prime}_{\mu \nu}{f^{(-)}}^{{\prime}\mu \nu}
\end{displaymath}
\be\label{165}
+ (d^{\prime}_{\mu} \xi^{\prime}_{\mu})^2 + \hbox{total derivative terms};
\ee

\be\label{166} \Gamma_{eff}^{(4)} = {N_c \over 16\pi^2} \Gamma(2,x)
tr\left[{1 \over 12}(1+\gamma_{21})\xi^{\prime}_{\mu}\xi^{\prime}_{\nu}
{\xi^{\prime}}^{\mu}{\xi^{\prime}}^{\nu}
- {1 \over 6}(1+\gamma_{22})\xi^{\prime}_{\mu}{\xi^{\prime}}^{\mu}
\xi^{\prime}_{\nu}{\xi^{\prime}}^{\nu}\right].\ee

\noindent
In the presence of gluonic interactions, there appear then 10 unknown
constants:
$\gamma_{-1}$; $\gamma_{01}$, $\gamma_{02}$, $\gamma_{03}$; $\gamma_{11}$,
 $\gamma_{12}$,
$\gamma_{13}$, $\gamma_{14}$; $\gamma_{21}$, $\gamma_{22}$. To these, we have
to
 add the
original  $G_S$ and $G_V$ constants, as well as the scale $\Lambda_{\chi}$.
 However, as
already mentioned in section 4,
the unknown constant $(1+\gamma_{-1})$ in eq. (\ref{162})
can be traded by an appropriate change of the scale $\Lambda_{\chi}$,
\be\label{167}
\Gamma(-1,\tilde x)=\Gamma(-1,x)\{1+\gamma_{-1}\}\ ;\ \tilde
x={M^2_Q\over\tilde{\Lambda}^2_{\chi}}, \ee
and a renormalization of the constant $G_S$,
\be\label{168}
G_S\to \tilde G_S={\tilde{\Lambda}^2_{\chi}\over\Lambda^2_{\chi}}G_S\ .
\ee

\noindent Altogether, we then have 12 (a priori unknown) theoretical constants
and one scale $\Lambda_{\chi}$. They determine 18 non-trivial physical
couplings
(in the large $N_c$-limit)
of the low energy QCD effective Lagrangian:

 $<\bar{\psi} \psi>$, $f_{\pi}$, $L_1$, $L_3$, $L_5$, $L_8$,
$L_9$, $L_{10}$, $H_1$, $H_2$, $f_V$, $f_A$, $g_V$, $c_m$, $c_d$, $M_S$, $M_V$
and $M_A$.

In full generality, the results are a follows:
\be\label{170}
<\bar {\psi} \psi> =
-{N_c \over  16 \pi^2} 4\,M_Q^3 \Gamma_{-1}(1 + \gamma_{-1}).
\ee

\be\label{171}
{1\over 4}f_{\pi}^2 = {N_c\over 16\pi^2} M_Q^2 g_A\Gamma_0(1 + \gamma_{01}).
\ee
\vskip 1cm
\be\label{172} L_2 = 2L_1 = {N_c \over 16\pi^2}{1\over 24}\times
\ee
\begin{displaymath}
\left[(1-g_A^2)^2\Gamma_0(1 + \gamma_{03})
+ 4g_A^2 (1-g_A^2) \Gamma_1(1 + {3 \over 2}\gamma_{12}-{1 \over 2}\gamma_{13})
+ 2g_A^4 \Gamma_2(1 + \gamma_{21})\right].\end{displaymath}

\begin{displaymath}
 \tilde L_3 = {N_c \over 16\pi^2}{1\over 24}\left[-3(1-g_A^2)^2
\Gamma_0(1 + \gamma_{03}) + 4g_A^4 \Gamma_1(1 + \gamma_{13})\right.
\end{displaymath}
\be\label{173}
\left.- 12 g_A^2 (1-g_A^2)\Gamma_1(1 + {3 \over 2}\gamma_{12}-{1 \over
 2}\gamma_{13})
- 8g_A^4\Gamma_2(1 + {1\over 2}(\gamma_{21}+\gamma_{22})),
\right]
\ee
\be\label{174}
L^S_3={c^2_d\over 2M^2_S}.
\ee

\be\label{175}
L_5 = {N_c\over 16\pi^2} {1\over 4} g_A^3 {1 + \gamma_{01}\over 1 +
\gamma_{02}}
\left[\Gamma_0(1 + \gamma_{01})-\Gamma_1(1 + \gamma_{11})\right].
\ee

\be\label{177}
L_8 = {N_c \over 16\pi^2} \left[{1\over 16}{1 + \gamma_{01}\over 1 +
 \gamma_{02}}
-{1\over 24}{\Gamma_1(1 + \gamma_{13})\over \Gamma_0(1 + \gamma_{01})}\right]
g_A^2\Gamma_0(1 + \gamma_{01}).
\ee
\be\label{179}
L_9 = {N_c \over 16\pi^2}{1\over 6}\left[(1-g_A^2)
\Gamma_0(1 + \gamma_{03}) +
2 g_A^2\Gamma_1(1 + {3 \over 2}\gamma_{12}-{1 \over 2}\gamma_{13})\right].\ee
\be\label{180}
L_{10} = -{N_c \over16\pi^2}{1\over6}\left[(1-g_A^2)
\Gamma_0(1 + \gamma_{03})+
g_A^2\Gamma_1(1 + \gamma_{13})\right].\ee
\be\label{181} H_1 = -{N_c \over16\pi^2}{1\over12}\left[(1+g_A^2)
\Gamma_0(1 + \gamma_{03}) -
g_A^2\Gamma_1(1 + \gamma_{13})\right].\ee
\be\label{182}
\tilde H_2 = {N_c \over16\pi^2}{1 \over 12}\left[6 \rho^2
\Gamma_{-1}(1 + \gamma_{-1}) -
6\rho^2 \Gamma_0(1 + \gamma_{02}) -
6\rho g_A \Gamma_0(1 + \gamma_{01}) +
g_A^2\Gamma_1(1 + \gamma_{13})\right],
\ee
\be\label{182bis}
H_2^S={c_m^2\over M^2_S}.
\ee
\vskip 1cm
\be\label{183}
f_V = \sqrt 2 \lambda_V
\ee
and
\be\label{184}
f_A = \sqrt 2 g_A \lambda_A,
\ee
\noindent with
\be\label{185}
\lambda_V^2 = {N_c \over 16\pi^2} {1 \over 3}\Gamma_0(1 + \gamma_{03});
\ee
\noindent and
\be\label{186}
\lambda_A^2 ={N_c \over 16\pi^2} {1 \over 3}
\left[\Gamma_0(1 + \gamma_{03})-
\Gamma_1(1 + \gamma_{13})\right].
\ee
\be\label{187}
g_V = {N_c \over 16\pi^2} {1 \over \lambda_V} {\sqrt 2 \over 6}
\left[(1-g_A^2) \Gamma_0(1 + \gamma_{03}) +
2 g_A^2 \Gamma_1(1 + {3 \over 2}\gamma_{12}-{1 \over 2}\gamma_{13})\right].
\ee
\be\label{188}
c_m = {N_c \over 16\pi^2} {M_Q \over \lambda_S} \rho
\left[\Gamma_{-1}(1 + \gamma_{-1}) -
2 \Gamma_0(1 + \gamma_{02})\right],
\ee
\be\label{189}
c_d = {N_c \over 16\pi^2} {M_Q \over \lambda_S} 2 g_A^2\left[
\Gamma_0(1 + \gamma_{01})
-\Gamma_1(1 + \gamma_{11})\right],
\ee
\noindent with
\be\label{190}
\lambda_S^2 = {N_c \over 16\pi^2} {2 \over 3} \left[3
\Gamma_0(1 + \gamma_{01})
- 2\Gamma_1(1 + \gamma_{14}) \right].
\ee
\vskip 1cm
\be\label{191}
M_S^2 = {N_c \over 16\pi^2} {8M_Q^2  \over \lambda_S^2}
\Gamma_0(1 + \gamma_{02}).
\ee
\be\label{192}
M_V^2 = {3 \over 2} {\Lambda_{\chi}^2 \over G_V(\Lambda_{\chi}^2)}
{1 \over \Gamma(0,x)(1 + \gamma_{03})}.
\ee
\be\label{193}
M_A^2\left\{1- {\Gamma(1,x)(1 + \gamma_{13}) \over
\Gamma(0,x)(1 + \gamma_{03})}\right\} = M_V^2 +
6 M_Q^2{1 + \gamma_{01} \over 1 + \gamma_{03}}.
\ee

There exist  relations among the physical couplings above which are independent
of the unknown gluonic constants. They are clean tests of the basic
assumption that the low energy effective action of QCD follows from an ENJL
Lagrangian of the type considered here. The relations are
\be\label{194}
f_V^2M_V^2 - f_A^2M_A^2 = f_{\pi}^2\,\,\, (1^{st}\,\hbox{ Weinberg sum rule}),
\ee

\be\label{195}
L_9={1\over 2}f_V g_V\ ,
\ee
\be\label{196}
L_{10}=-{1\over 4}f_V^2 + {1\over 4}f_A^2\ ,
\ee
\be\label{197}
2H_1=-{1\over 4}f_V^2 - {1\over 4}f_A^2\ ,
\ee
\noindent and
\be\label{198}
{H_2+2L_8\over 2L_5}={c_m\over c_d}\ .
\ee

The first four relations have already been discussed in the previous section.
The combination of couplings in the
r.h.s. of eq.(\ref{198}) is the one which
 appears in the context
of non leptonic weak interactions, when one considers weak decays like $K\to
\pi
 H$ (light Higgs) \cite{52}.
In fact, from the low energy theorem derived in \cite{33}
it follows that
\be\label{199}
{H_2+2L_8\over 2L_5}={1\over 4}{{<0|\bar ss|0>\over <0|\bar uu|0>}-1\over
 f_K/f_{\pi}-1}\ .
\ee

\noindent Experimentally
\be\label{200}
f_K/f_{\pi}-1=0.22\pm 0.01\ .
\ee

\noindent Unfortunately, the numerator in the r.h.s. of (\ref{199})
is poorly known. If
we vary the ratio
\be\label{201}
{<\bar ss>\over <\bar uu>}-1\quad\hbox{ from }\ -0.1\ \hbox{ to }\ -0.2\ ,
\ee
\noindent as suggested by the authors of ref. \cite{52}, then
eq.(\ref{198}) leads to the estimate
\be\label{202}
c_m/c_d=-1.1\times 10^{-1}\ \hbox{ to }\ -2.3\times 10^{-1}\ .
\ee
\noindent With this estimate incorporated in eq.(\ref{53}), we are led to the
 conclusion that
\be\label{203}
|c_d|\simeq 34 MeV\ .
\ee
\noindent In the version corresponding to the $1^{st}$ alternative,
the results for $c_m$ and $c_d$ are those in
eqs.(\ref{127}) and (\ref{128}). We observe that in this case $c_m/c_d$
comes out always positive for reasonable values of $M^2_Q/\Lambda^2_{\chi}$.

\subsection{Gluonic correction to $O(\alpha_S N_c)$.}

\quad We can make an estimate of the ten constants
$\gamma_{-1};$
$\gamma_{01},\gamma_{02},\gamma_{03};$ $\gamma_{11},\gamma_{12},
\gamma_{13},\gamma_{14};$ $\gamma_{21}$ and $\gamma_{22}$
by keeping only the leading
 contribution
which involves the gluon vacuum condensate ${<{\alpha_S
\over \pi} GG> \over M_Q^4}$ as was done in Ref. \cite{13}. The relevant
dimensionless parameter is

\be\label{204}
g = {\pi^2 \over 6N_c}{<{\alpha_S \over \pi} GG> \over M_Q^4}.
\ee

\noindent Notice that in the large-$N_c$ limit $g$ is a parameter of $O(1)$.
One should also keep in mind that the gluon average in (\ref{204})
is the one corresponding to fluctuations below
the $\Lambda_{\chi}$-scale. The relation
of $g$ to the conventional
gluon condensate which appears in the QCD sum rules \cite{42}
is rather unclear. We are forced to consider $g$ as a free parameter.
Up to order $O(\alpha_S N_c)$,
this is the only unknown quantity which appears,
and we can express all the $\gamma$'s in terms of $g$. We find:

\be\label{205}
\gamma_{-1} = {\Gamma(1,x) \over \Gamma(-1,x)}\,2\,g;
\ee

\be\label{206}
\pmatrix{\gamma_{01}\cr \gamma_{02}\cr \gamma_{03}\cr}
= {\Gamma(2,x) \over \Gamma(0,x)}\pmatrix{1\cr 2\cr -3/5\cr}g;
\ee

\be\label{207}
\pmatrix{\gamma_{11}\cr \gamma_{12}\cr \gamma_{13}\cr \gamma_{14}\cr}
= {\Gamma(3,x) \over \Gamma(1,x)}\pmatrix{1\cr 1/5\cr 3/5\cr 9/5\cr}g;
\ee

\be\label{208}
\pmatrix{\gamma_{21}\cr \gamma_{22}\cr}
= {\Gamma(4,x) \over \Gamma(2,x)}\pmatrix{0\cr 2/5\cr}g.
\ee

\noindent Notice that the combination ${3 \over 2}\gamma_{12} -
{1 \over 2}\gamma_{13}$ entering some of the $L_i$'s coupling constants
is zero. This is the reason why in Ref. \cite{13}
it was found that in the limit $g_A\to
 1$, $L_2$ and $L_9$
have no gluon correction of $0(\alpha_sN_c)$.

To this approximation, we have then reduced the theoretical
parameters to three unknown constants

\begin{displaymath}
G_S,\,\, G_V\,\, \hbox{and}\,\, g;
\end{displaymath}

\noindent and the scale $\Lambda_{\chi}$.

\section{COMPARISON WITH OTHER MODELS.}

\quad It is interesting that practically all the models of low energy QCD
 discussed in
 the literature can
be obtained as some limit of the ENJL model developed in
 the previous
sections. Here, we wish to show this for some of them.

As already mentioned, the QCD effective action approach
proposed in Ref. \cite{13} is obtained in the limit where
\begin{displaymath}
G_V\to 0 \,\,\, \hbox{and} \,\,\, <H> = M_Q
\end{displaymath}

\noindent Only $M_Q$ and $\Lambda_{\chi}$, which in Ref. \cite{13}
is taken as an
 ultraviolet cut-off, with
$M_Q/\Lambda_{\chi}\to 0$ whenever the limit exists, are left as basic
 parameters. In this case $g_A = 1$.

The comparison with the model of Manohar and Georgi \cite{39}
is more subtle. These  authors suggest as
an effective Lagrangian of QCD at energies
below the chiral symmetry breaking  scale, the Lagrangian
(in our notation) :
\begin{displaymath}
{\cal L}_{G-M}=i\bar Q\gamma^{\mu}(\partial_{\mu}+iG_{\mu}+\Gamma_{\mu})Q+
{i\over 2}g_A\bar Q \gamma_5\gamma^{\mu}\xi_{\mu}Q-M_Q\bar QQ
\end{displaymath}
\be\label{209}
+{1\over 4}f^2tr \partial_{\mu}U \partial^{\mu}U-{1\over
4}tr\sum_{a}^{}G_{\mu \nu}^{(a)}G^{(a) \mu \nu}+\cdots\ ,
\ee
\noindent Couplings to external fields are not discussed in \cite{39}.
The dots in (\ref{209})
 stand for terms of
higher chiral dimension which in practice are ignored. In the limit $g_A=1,f\to
 0$ and with the
external fields $l_{\mu}=r_{\mu}=0$,
this Lagrangian coincides with the one
 proposed in Ref. \cite{13}.
We have already discussed how the mass
term $M_Q\bar QQ$ originates in the Nambu  Jona-Lasinio
approach. What is the origin of the explicit
kinetic term of Goldstone fields above; i.e., the
coupling $f$? As we have seen in section 5,
the integration over constituent  quarks in a
Lagrangian like the one above produces
such a kinetic term with the result as  shown in
eq.(\ref{124}) (with gluons ignored).
Remember also that integration of the vector
and axial-vector  mesons, as well as the
scalars in the Nambu Jona-Lasinio approach did not contribute to the
 kinetic term of Goldstone fields.
On the other hand the $g_A$-coupling, with $g_A<1$, appears naturally in
 the Nambu
Jona-Lasinio approach. It originates in the mixing of the primitive $G_S$
and $G_V$ four-fermion
couplings; and the ``prediction'' for $g_A$ is
\be\label{210}
g_A=1-{f^2_{\pi}\over f^2_VM^2_V}\ .
\ee

\noindent In conclusion we find that the Lagrangian of Georgi and Manohar can
be
 understood within the
framework of a Nambu Jona-Lasinio mechanism, provided $f$ in (\ref{209})
is very small
 (perhaps coming from
having integrated out the baryon's degrees of freedom).

Models like the one presented in Ref. (\cite{51})
are less well defined and therefore more
 difficult to compare with.
They have to do more with what we call the $1^{st}$ alternative
 and the way to reach
a local approximation to the ''gluonless'' effective action.

The hidden gauge vector meson model of Bando, Kugo and Yamawaki \cite{43}
is a chiral effective
Lagrangian of vector fields coupled
to Goldstone fields which, as discussed in \cite{37}, automatically
satisfies the QCD short-distance constraints necessary
to remove ambiguities
 when the vector
fields are integrated out. The model however has extra symmetries, like
\be\label{211}
g_V={1\over 2}f_V\ ,
\ee
\noindent and
\be\label{212}
L_2=2L_1={1\over 4}g^2_V
\ee
\noindent and
\be\label{213}
L_3=-3L_2
\ee
\noindent which are not automatically implemented in the Nambu Jona-Lasinio
 approach that we have
been discussing. As we shall see however,
our numerical predictions satisfy rather closely these relations.

The most economical phenomenological effective chiral Lagrangian with vector
and
 axial-vector fields
is the one proposed in Ref. \cite{37}, where
\be\label{214}
  f_V=\sqrt 2{f_{\pi}\over M_V}\ ;\quad
  g_V={1\over \sqrt 2}{f_{\pi}\over M_V}\ ;\quad
  f_A={f_{\pi}\over M_A}\ ;\quad
  M_A=\sqrt 2 M_V\ .
\ee

\noindent It predicts the five $0(p^4)$-couplings which exist in the chiral
 limit as follows:
\be\label{215}
L_3=-3L_2=-6L_1=-{3\over 4}L_9=L_{10}=-{3\over 8}{f_{\pi}^2\over M^2_V}\ .
\ee
 It also satisfies the first and second Weinberg sum rules :
\be\label{216}
f^2_{\pi}+f^2_AM^2_A=f^2_VM^2_V
\ee
\noindent and
\be\label{217}
f^2_VM^4_V=f^2_AM^4_A\ .
\ee

{}From eqs.(\ref{172}) to (\ref{180}) it appears that a very similar result
 emerges in the limit where
$g_A\to 0$: \be\label{218}
L_3=-3L_2=-6L_1=-{3\over 4}L_9={3\over 4}L_{10}=-{3\over 16} f^2_V\ .
\ee
In this limit, we also find  the relation
\be\label{219}
g_V={1\over 2}f_V\ .
\ee
These are in fact the relations derived in Ref. \cite{37} when the
contribution from the axial vector mesons is removed.

As we have seen in sections 5 and 6 the first Weinberg sum rule also follows
 from the Nambu
Jona-Lasinio model we have been considering. The second Weinberg sum rule,
as given in eq. (\ref{217}), is however only satisfied
if one arbitrarily requires
\be\label{220}
g_A={M^2_V\over M^2_A}
\ee
or, equivalently,
\be\label{221}
g^2_A={f^2_A\over f^2_V}\ .
\ee

\section{DISCUSSION OF NUMERICAL RESULTS.}

\quad In the ENJL model, we have three input parameters:

\be\label{222}
G_S\,\,,\,\,G_V\,\,\hbox{and}\,\,\Lambda_{\chi}.
\ee

\noindent The gap equation introduces a constituent chiral
quark mass parameter  $M_Q$, and the ratio

\be\label{223}
x = {M_Q^2 \over \Lambda_{\chi}^2}
\ee

\noindent is constrained to satisfy the equation

\be\label{224}
{1\over G_S} = x \Gamma(-1,x)(1+\gamma_{-1}).
\ee

\noindent Once $x$ is fixed, the constants $g_A$ and $G_V$ are related by
the equation

\be\label{225}
g_A = {1 \over 1 + 4 G_V x \Gamma(0,x)(1+\gamma_{01})}.
\ee

\noindent Therefore, we can trade $G_S$ and $G_V$ by
$x$ and $g_A$;
but we need an observable to fix the scale $\Lambda_{\chi}$. This
 is the scale
which determines the $\rho$ mass in eq. (\ref{192}), i.e.,

\be\label{226}
\Lambda_{\chi}^2 = {2 \over 3} M_V^2 G_V \Gamma(0,{M_Q^2 \over
 \Lambda_{\chi}^2})(1+\gamma_{03}).
\ee

There are various ways one can proceed. We find it useful to fix as input
 variables the values of
$M_Q$, $\Lambda_{\chi}$ and $g_A$. Then we have predictions for

\be\label{227}
f_{\pi}^2\,\,\, , \,\,\, <\bar{\psi} \psi>
\ee
\be\label{228}
M_S\,\,\, , \,\,\, M_V \,\,\, \hbox{and} \,\,\, M_A
\ee
\be\label{229}
f_V\,\, , \,\,g_V\,\, \hbox{and} \,\,f_A\,\,\,;\,\,\, \hbox{and}
\,\,\,c_m\,\,\,
 \hbox{and} \,\,\,c_d
\ee

\noindent and the $O(p^4)$ couplings:

\be\label{230}
L_i\,\,\,i=1,2,\ldots,10\,\,\, \hbox{and} \,\,\,H_1\,\,\, , \,\,\,H_2.
\ee

\noindent In principle we can
also calculate any higher $O(p^6)$ coupling which
 may become of
interest. So far, we have fixed twenty-two parameters.
Eighteen of them are
 experimentally known.

In the first column of Table \ref{table2}
we have listed the experimental values of
the parameters which we consider.
In comparing with the predictions of the ENJL model
it should be kept in mind that the relations (\ref{194}) to (\ref{196})
are satisfied by the model while relations (\ref{211}) to (\ref{213})
only have numerically small corrections. These relations are rather well
satisfied by the experimental values and thus constitute a large part
of the numerical success of the model.

We have also used the predictions leading in $1/N_c$ so we have
$ L_1 = L_2 /2$, $L_4 =  L_6 = 0$
and we do not consider $L_7$ since this is
given mainly by the $\eta^\prime$-contribution \cite{33}.
In evaluating the  predictions given in Table \ref{table2}
we have used the full expressions for the incomplete gamma functions
and the numerical value of the $\gamma_{ij}$ in terms of g given in
eqs. (\ref{205}) to (\ref{208}).

The first column of errors in Table \ref{table2}
shows the experimental ones. The second column
gives the errors we have used for the fits. When no error is indicated in this
column, it means
that we never use the corresponding parameter for fitting. This is the
case for $<\overline{q}q>$ which is quadratically divergent in the
cut-off and which is not very well known experimentally. This is also the
case for $c_m$ which depends on $<\overline{q}q>$.
Fit 1 corresponds to a least squares fit with the maximal set of parameters
and requiring $g \ge 0$. Fit 2 corresponds to a fit
where only $f_\pi$ and the $L_i$ are used as input in the fit while
fit 3 has as additional input the vector and scalar mass.
The next column, fit 4, is the one where we require $g_A = 1$; i.e.,
we start with a model without the vector four-quark interaction. Here
there are no explicit
vector (axial) degrees of freedom so those have been dropped in this case.
This fit includes
all parameters except $M_V$, $M_A$, $f_V$, $g_V$
and $f_A$. Finally, fit 5 is the fit to all data keeping the gluonic
parameter $g$ fixed at a value of 0.5. The main difference with fit 1 is
a decrease in the value of $M_Q$. The value of $\Lambda_\chi$
changes very little.

The typical variation of the results with respect to
each of the 4 input parameters
can be judged from the results in table \ref{table3}. Here we have
changed each of the input parameters by the error quoted by the
least squares fitting routine (MINUIT). Typically, this changes the overall
fit by about one standard deviation with respect to the input errors
quoted in table  \ref{table2}
(third column). The results of fit 2 were used as the standard.
In this fit
it was the value of $f_\pi$ and $L_9$ that constrained
the input parameters most.

The expected value for the parameter $g$ if we take typical values
from, e.g, QCD sum rules is of $O(1)$. None of the fits here really
makes a qualitative difference between a $g$ of about $0.5$ to $0$.
Numerically we can thus not decide between the two alternatives mentioned
in the introduction. This can be easily seen by comparing fit 1 and fit 5
in table \ref{table2}.

In all cases acceptable predictions for all relevant parameters are
possible. The scalar sector parameters tend all to be a bit on the low
side; but so is the constituent quark mass. The
predictions for the $L_i$'s are reasonably stable versus a variation of
the input parameters.
For $L_5$ and $L_8$ this is a major improvement as compared to the predictions
of the mean field approximation \cite{13}.

\section{CONCLUSIONS.}

\quad In this paper we have determined the low energy effective action of
an ENJL cut-off version of QCD to $O(p^4)$.
We have done this in two alternative scenarios. One where
the ENJL model is a local approximation to the QCD effective action
where the gluons have been fully integrated over. In the other
scenario we envisaged the four-fermion operators as the result of
integrating over both quark and gluon
degrees of freedom down to a scale $\Lambda_{\chi}$,
and then performing a local expansion
of that effective action. In this way we have avoided possible
problems of double-counting of contributions. For example, the value
of $L_9$ can be well reproduced within a simple quark model
approach \cite{13}; and also as coming from integrating out
the vector mesons \cite{36}. By studying this within the
context of a well defined model where both degrees of freedom
are present we have clarified this ambiguity.

The ENJL model incorporates in various limits most of the effective
low-energy models discussed in the literature.
In particular the gauged Yang-Mills vector Lagrangian \cite{43},
the Georgi-Manohar effective quark-meson model \cite{39} and the
QCD effective action approach model \cite{13}. It also incorporates
most of the short distance relations which are expected in
QCD \cite{37}. The
derivation of these relations has been done including all possible
gluonic interactions to leading order in the $1/N_c$-expansion.
The relations are valid for all
models where the effective quark meson Lagrangian can be written in
the form of eq. (\ref{105}). This is one of the main results of this paper.

The other short distance relations derived using the Froissart bound
and the second Weinberg sum rule \cite{37} are not explicitly satisfied
by the model. The deviations are however suppressed by
terms of order $O(g_A^4)$, so they can be
recovered in an appropriate limit. For reasonable values of the
input parameters these deviations are very small.

The numerical predictions in terms of only 4 input parameters are very
successful. The constituent quark mass is, however, somewhat smaller
than expected. Both alternatives turn out to work rather well. It is
only when the scalar mass is required to be close to its measured value
that gluonic corrections become important.

In conclusion, the extended Nambu Jona-Lasinio model incorporates
a surprisingly large amount of information of the short distance
structure of QCD. This is probably the main reason for the successful
predictions of this and related models.

\vskip 1cm

{\bf ACKNOWLEDGMENTS}

We would like to thank C.H. Llewellyn Smith, A. Pich, J. Prades
and J. Taron
for useful conversations. We also thank L. H\"ornfeldt for
help in using his
program STENSOR.

\newpage
\begin{table}[htb]
\caption{Experimental values and predictions of the ENJL model
for the various low energy parameters discussed in the text.
All dimensionful quantities are in MeV. The difference
between the predictions is explained in the text.}
\label{table2}
\begin{center}
\begin{tabular}{|c|c|c|c|c|c|c|c|c|}
\hline
    & exp. & exp.  & fit   & fit 1 & fit 2 & fit 3 & fit 4 & fit 5 \\
    & value& error & error &       &       &       &       &       \\
\hline
\hline
$f_\pi$ & 86(${}^{\dagger}$) &  $-$  & 10     &  89  &  86  & 86    &  87&83\\
\hline
$\sqrt[3]{-<\overline{q}q>}$&194(${}^{\#}$)&8(${}^{\#}$)&$-$&281& 260 &  255
& 178 & 254  \\
\hline
$ 10^3 \cdot L_2$ & 1.2  &  0.4 &0.5  & 1.7   &  1.6 & 1.6 & 1.6 &1.7\\
$ 10^3 \cdot L_3$ &$-3.6$&  1.3 &1.3  & $-4.2$&$-4.1$&
$-4.4$&$-5.3$&$-4.7$\\
$ 10^3 \cdot L_5$ &  1.4 &0.5   & 0.5 & 1.6   &  1.5 &  1.1    &1.7&1.6\\
$ 10^3 \cdot L_8$ &  0.9 & 0.3  & 0.5 & 0.8   &  0.8 &  0.7    &1.1&1.0\\
$ 10^3 \cdot L_9$ &  6.9 &  0.7 &0.7  & 7.1   &  6.7 &  6.6   &5.8&7.1\\
$ 10^3 \cdot L_{10}$&$-5.5$& 0.7 & 0.7&$-5.9$ &$-5.5$&$-5.8$&
$-5.1$&$-6.6$\\
$ 10^3 \cdot H_1$ & $-$  &$-$&$-$  & $-4.7$ &$-4.4$&$-4.0$&$-2.4$&
$-4.6$\\
$ 10^3 \cdot H_2$ & $-$  &$-$&$-$  & $-0.3$ &$-0.3$&$-0.4$  &
 $-0.5$&$-0.8$\\
\hline
$ M_V $&768.3 &0.5& 100 &    811  &  830 & 831     &  $-$ & 802\\
$ M_A $& 1260&30&300 &     1331  & 1376 & 1609    &  $-$  & 1610\\
$ f_V $& 0.20&(*)& 0.02&   0.18  & 0.17 & 0.17    &  $-$  & 0.18\\
$ g_V $&0.090&(*)& 0.009&    0.081  & 0.079&  0.079  & $-$ & 0.080\\
$ f_A $&0.097&0.022(*)& 0.022&   0.083  & 0.080& 0.068   & $-$&0.072\\
\hline
$ M_S $&983.3&2.6&200&            617  & 620  &  709    & 989 & 657\\
$ c_m $& $-$&$-$& $-$ &           20  &  18  &  20     & 24  & 25\\
$ c_d $&  34 &(*)&  10&          21  &  21  &  18     & 23   & 19\\
\hline
\hline
$x$  & & & &              0.052  & 0.063 &  0.057 &  0.089 & 0.035\\
$g_A$& & & &               0.61  &  0.62 &  0.62  &  1.0   & 0.66\\
$M_Q$& & & &                265  & 263   &  246   &  199   & 204 \\
$g$  & & & &                0.0  &  0.0  &  0.25  &  0.58  & 0.5 \\
\hline
\end{tabular}
\end{center}

(${}^{\dagger}$) This corresponds to $f_0$ which is the value of $f_{\pi}$
in the chiral limit.

(${}^{\#}$) See Ref. \cite{53}.

(*) In addition to the experimental error, the chiral loop corrections to
these parameters have not been calculated.

\end{table}

\begin{table}[htb]
\caption{Experimental values and predictions for these quantities in
the ENJL model. The input parameters are chosen such that the typical
variation with input parameters can be seen. Set 1 is the standard,
corresponding to fit 2 in Table \protect{\ref{table2}}.}
\label{table3}
\begin{center}
\begin{tabular}{|c|c|c|c|c|c|c|}
\hline
    & exp. & set 1 & set 2 & set 3 & set 4 & set 5 \\
\hline
\hline
$f_\pi$ & 86 &                   86    &  81  &  91     &   98 &97 \\
\hline
$\sqrt[3]{-<\overline{q}q>}$&194&260& 236 &  260     & 296&267     \\
\hline
$ 10^3 \cdot L_2$ &  1.2       & 1.6   &  1.5 & 1.6     & 1.6  &1.5  \\
$ 10^3 \cdot L_3$ &$-3.6$      & $-4.1$&$-4.0$&$-3.6$   &$-4.1$&$-4.5$ \\
$ 10^3 \cdot L_5$ &  1.4       & 1.5   &  1.3 &  2.1    &  1.5 &0.7  \\
$ 10^3 \cdot L_8$ &  0.9       & 0.8   &  0.7 &  0.9    &  0.7 &0.6  \\
$ 10^3 \cdot L_9$ &  6.9       & 6.7   &  6.2 &  6.6   &   6.7 &6.0 \\
$ 10^3 \cdot L_{10}$&$-5.5$    &$-5.5$ &$-5.1$&$-5.1$   & $-5.5$&$-5.7$\\
$ 10^3 \cdot H_1$ &  $-$      & $-4.4$ &$-3.9$&$-4.6$  & $-4.3$&$-3.1$ \\
$ 10^3 \cdot H_2$ & $-$       & $-0.3$ &$ 0.4$&$ 0.7$  & $ 0.5$&0.1 \\
\hline
$ M_V $&768.3 &                     830  &  828 &  984    &  945  &1016\\
$ M_A $& 1260&                   1376  & 1420 & 1540    &  1567 &3662 \\
$ f_V $& 0.20&                   0.17  & 0.16 & 0.17    &  0.17 &0.15 \\
$ g_V $&0.090&                  0.079  & 0.077&  0.078  &  0.079&0.077 \\
$ f_A $&0.097&                  0.080  & 0.074& 0.090   &  0.080&0.034 \\
\hline
$ M_S $&983.3&                      620  & 630  &  620    & 706   &1102\\
$ c_m $& $-$ &                     18  &  16  &  20     & 21  &26   \\
$ c_d $&  34 &                     21  &  18  &  26     & 23  &16   \\
\hline
\hline
$x$& &                          0.063  & 0.08  &        &      &    \\
$g_A$& &                         0.62  &       &  0.7   &      &   \\
$M_Q$& &                          263  &       &        &  300 &   \\
$g$ & &                           0.0  &       &        &      &0.6 \\
\hline
\end{tabular}
\end{center}
\end{table}

\vspace*{10cm}

\newpage
\appendix
{\bf APPENDIX}

\section{Euclidean conventions.}
To go to Euclidean space, we adopt the usual prescription

\be\label{A1}
x^{\mu} \equiv (x^0,x^i) = (-i\bar{x}^0, \bar{x}^i)
\ee

\be\label{A2}
\partial_{\mu} \equiv (\partial_0,\partial_i) =
(+i\bar{\partial}_0, \bar{\partial}_i)
\ee

\noindent For arbitrary $4$-vectors $a_{\mu},b_{\mu}$, one has then

\be\label{A3}
a^{\mu}b_{\mu} = - \bar{a}_{\mu}\bar{b}_{\mu}.
\ee

\noindent It is convenient to work with hermitian gamma matrices with
positive metric. We take

\be\label{A4}
\tilde{\gamma}_{\mu} \equiv - i\bar{\gamma}_{\mu} ,
\ee

\noindent i.e., $\tilde{\gamma}_0 = -\gamma_o$ and $\tilde{\gamma}_i =
-i\gamma_i$, which have the required properties

\be\label{A5}
\tilde{\gamma}_{\mu}^+ = \tilde{\gamma}_{\mu} \qquad ; \qquad
\{\tilde{\gamma}_{\mu}, \tilde{\gamma}_{\mu}\} = 2\delta_{\mu\nu} .
\ee

\noindent Other useful relations are

\be\label{A6}
\gamma_5 \equiv -i\gamma_0\gamma_1\gamma_2\gamma_3 = \bar{\gamma}_0
\bar{\gamma}_1\bar{\gamma}_2\bar{\gamma}_3 = \tilde{\gamma}_0\tilde{\gamma}_1
\tilde{\gamma}_2\tilde{\gamma}_3 = \gamma_5^{\dagger}
\ee

\be\label{A7}
\bar{\sigma}_{\mu\nu} = {i\over 2} [\bar{\gamma}_{\mu},
\bar{\gamma}_{\nu}] = - {i\over 2} [\tilde{\gamma}_{\mu},
\tilde{\gamma}_{\nu}] .
\ee

\noindent In the following of this appendix we will omit the bars in the
Euclidean quantities.

\section{Covariant derivatives.}
We shall next recall the Seeley-DeWitt coefficients of the Heat kernel
expansion of the operator

\be\label{A8}
{\cal D}^{\dagger}_E {\cal D}_E - M^2_Q \equiv -
\nabla_{\mu}\nabla_{\mu} + E,
\ee

\noindent with $\nabla_{\mu}$ the full covariant derivative in eq. (\ref{100})
reexpressed in terms of the ``primed'' external fields,
\be \label{A9}
\nabla_{\mu}=\partial_{\mu} + iG_{\mu} + \Gamma_{\mu}^{\prime} - {i \over 2}
 \gamma_5
\xi_{\mu}^{\prime} , \ee
\noindent ${\cal D}_E$ the euclidean Dirac operator defined in eq. (\ref{105}),
and $E$ the quantity \cite{13} :

\begin{displaymath}
E =  {i \over 4} \widetilde\gamma_{\mu} \gamma_5
\{\xi_{\mu}^{\prime},\Sigma^{\prime}- \gamma_5 \Delta^{\prime}\}
- {i\over 2}\sigma_{\mu\nu}
R^{\prime}_{\mu\nu}+ {1\over 4}{\Sigma^{\prime}}^2 - M_Q^2
\end{displaymath}
\be\label{A10}
- {1\over 4}{\Delta^{\prime}}^2
- {1 \over 8} \gamma_5 [\Sigma^{\prime},\Delta^{\prime}]
+ {1\over 2}\widetilde\gamma_{\mu}d^{\prime}_{\mu}(\Sigma^{\prime} -
\gamma_5 \Delta^{\prime}).
\ee

\noindent We recall that

\be \label{A11}
\Gamma_{\mu}^{\prime} = {1 \over
2}\{\xi_{\dagger}[\partial_{\mu}-ir_{\mu}^{\prime}]\xi +
\xi[\partial_{\mu}-il_{\mu}^{\prime}]\xi^{\dagger}\}
=\Gamma_{\mu} - {i\over 2}W^{(+)}_{\mu},\ee
\be\label{A12}
\xi_{\mu}^{\prime} =i\{\xi^{\dagger}[\partial_{\mu}-ir_{\mu}^{\prime}]\xi -
\xi[\partial_{\mu}-il_{\mu}^{\prime}]\xi^{\dagger}\}
=\xi_{\mu} - W^{(-)}_{\mu},\ee
\be\label{A13}
\Sigma^{\prime} = \xi^{\dagger} {\cal M^{\prime}} \xi^{\dagger} +
\xi {\cal M^{\prime}}^{\dagger} \xi
= \Sigma + 2 \sigma,\ee
\be\label{A14}
\Delta^{\prime} = \xi^{\dagger} {\cal M^{\prime}} \xi^{\dagger} -
\xi {\cal M^{\prime}}^{\dagger}\xi
= \Delta,\ee

\noindent where $l_{\mu}^{\prime}$, $r_{\mu}^{\prime}$ and $\cal{M^{\prime}}$
were defined in eqs. (\ref{102}) to (\ref{104}).

\noindent $d^{\prime}_{\mu}$ is the covariant derivative
with respect to the $\Gamma^{\prime}_{\mu}$-connection, i.e.,

\be\label{A15}
d^{\prime}_{\mu} A \equiv \partial_{\mu}A + [\Gamma^{\prime}_{\mu}, A],
\ee

\noindent $R^{\prime}_{\mu\nu}$ is the full strength
tensor, i.e.,

\be \label{A16}
R^{\prime}_{\mu \nu} = [\nabla_{\mu}, \nabla_{\nu}] =
i G_{\mu \nu} - {i\over 2}f^{\prime}_{\mu \nu},\ee

\noindent with

\be\label{A17}
G_{\mu\nu} = \partial_{\mu} G_{\nu} - \partial_{\nu} G_{\mu} +
i[G_{\mu},G_{\nu}],
\ee

\noindent and

\be\label{A18}
-{i \over 2}f_{\mu \nu}^{\prime} =  -{i \over 2} f_{\mu\nu}
- {1 \over 4}\gamma_5([\xi_{\mu},W_{\nu}]+[W_{\mu},\xi_{\nu}])
-{1 \over 4}[W_{\mu},W_{\nu}] - {i \over 2}W_{\mu\nu},\ee

\noindent where
\be \label{A19}
W_{\mu}=W^{(+)}_{\mu}-\gamma_5 W^{(-)}_{\mu},\ee
\be \label{A20}
f_{\mu\nu}=f^{(+)}_{\mu\nu}-\gamma_5 f^{(-)}_{\mu\nu},\ee
\be \label{A21}
W_{\mu\nu}=W^{(+)}_{\mu\nu}-\gamma_5 W^{(-)}_{\mu\nu}.\ee

\noindent Here
$f^{(\pm)}_{\mu\nu}$ and $W^{(\pm)}_{\mu\nu}$ are respectively defined as in
eqs. (\ref{44}) and (\ref{39}).

\section{Seeley-DeWitt coefficients.}
         The Heat kernel expansion of the operator $D^{\dagger}_E D_E - M^2_Q$
in eq. (\ref{A8}) defines the Seeley-DeWitt coefficients $H_n(x,x)$, which
were used in ref. \cite{13} up to $n=3$. Here we need them up to $n=6$. In the
literature, they have been computed up to $n=5$. Fortunately, $\gamma_{21}$ and
$\gamma_{22}$ at the order $\alpha_S N_c$ were calculated in Ref. \cite{13}
by other methods, and we have taken their results.
All the other quantities can be computed using only the $H_n$'s up to $n=5$.

In what follows, the $H_n$'s are defined up to a total derivative and a
circular permutation. See ref \cite{44} and \cite{45}.
Terms that do not contribute to our results are dropped.

\be\label{A22}
H_0(x,x) = 1,
\ee
\vskip 1cm
\be\label{A23}
H_1(x,x) = - E,
\ee
\vskip 1cm
\be\label{A24}
H_2(x,x) = {1\over 2} \big[E^2  +
{1\over 6} R_{\mu\nu}R_{\mu\nu}\big],
\ee
\vskip 1cm
\be\label{A25}
H_3(x,x) =  -{1\over 6} \big[E^3 - {1\over 2}E\nabla_{\mu}\nabla_{\mu}E
+ {1\over 2} ER_{\mu\nu}R_{\mu\nu}\big],
\ee
\vskip 1cm
\begin{displaymath}
H_4(x,x)  =
{1\over 24} \big[E^4 - E^2\nabla_{\mu}\nabla_{\mu}E
\end{displaymath}
\begin{displaymath}
+ {1\over 5} (ER_{\mu\nu}ER_{\mu\nu} + 4 E^2R_{\mu\nu}R_{\mu\nu})
-{4\over 15}\nabla_{\rho}\nabla_{\rho}ER_{\mu\nu}R_{\mu\nu}
\end{displaymath}
\begin{displaymath}
+ {17\over 210}R_{\mu\nu}R_{\mu\nu}R_{\rho\sigma}R_{\rho\sigma}
+ {2\over 35}R_{\mu\nu}R_{\nu\rho}R_{\mu\sigma}R_{\sigma\rho}
\end{displaymath}
\be\label{A26}
+ {1\over 105} R_{\mu\nu}R_{\nu\rho}R_{\rho\sigma}R_{\sigma\mu}
+ {1\over 420}R_{\mu\nu}R_{\rho\sigma}R_{\mu\nu}R_{\rho\sigma}\big],
\ee
\vskip 1cm
\begin{displaymath}
H_5(x,x) =
-{1\over 120} \big[E^5 - 2E^3\nabla_{\mu}\nabla_{\mu}E
\end{displaymath}
\begin{displaymath}
- E^2\nabla_{\mu}E\nabla_{\mu}E + {2\over 3} E^2R_{\mu\nu}ER_{\mu\nu}
+  E^3R_{\mu\nu}R_{\mu\nu}
\end{displaymath}
\begin{displaymath}
+ {2\over 3} ER_{\mu\nu}\nabla_{\mu}E\nabla_{\nu}E
- {8\over 7} E\nabla_{\rho}E\nabla_{\rho}ER_{\mu\nu}R_{\mu\nu}
- {4\over 21} R_{\mu\nu}E\nabla_{\rho}E\nabla_{\rho}ER_{\mu\nu}
\end{displaymath}
\begin{displaymath}
- {10\over 21} \nabla_{\mu}E\nabla_{\nu}ER_{\nu\rho}R_{\rho\mu}
+ {2\over 21} \nabla_{\mu}E\nabla_{\nu}ER_{\mu\rho}R_{\rho\nu}
\end{displaymath}
\be\label{A27}
+ {2\over 7} \nabla_{\mu}R_{\mu\nu}E\nabla_{\rho}ER_{\rho\nu}
- {11\over 21} \nabla_{\mu}E\nabla_{\mu}ER_{\nu\rho}R_{\nu\rho}
+ {1\over 42} \nabla_{\mu}R_{\nu\rho}E\nabla_{\mu}ER_{\nu\rho}\big].
\ee

\section{Effective actions.}

\label{appD}
\be \label{A28}
\Gamma_{eff}^{(1)} = {N_c \over
 16\pi^2}\Gamma(-1,x)
tr\left[2({1\over 4}{\Sigma^{\prime}}^2 - M_Q^2)\right],\ee

\begin{displaymath}
\Gamma_{eff}^{(2)} = {N_c \over 16\pi^2} \Gamma(0,x)
tr\left[{1\over 2}\{{\xi^{\prime}}^{\mu},\Sigma^{\prime}\}^2 +
{i\over 4}\{\xi^{\prime}_{\mu},\Sigma^{\prime}\}
{d^{\prime}}^{\mu}\Delta^{\prime}
- ({1\over 4}{\Sigma^{\prime}}^2 - M_Q^2)^2\right.
\end{displaymath}
\be\label{A29}
\left.+ {1\over 4}d^{\prime}_{\mu}\Sigma^{\prime}
{d^{\prime}}^{\mu}\Sigma^{\prime}
-{1\over 12}\left({f^{(+)}}^{\prime}_{\mu\nu}
{f^{(+)}}^{{\prime}\mu \nu}
+{f^{(-)}}^{\prime}_{\mu \nu}{f^{(-)}}^{{\prime}\mu \nu}\right)\right],
\ee

\begin{displaymath}
 \Gamma_{eff}^{(3)} = {N_c \over 16\pi^2} \Gamma(1,x)
 tr\left[
-{1 \over 16}({1\over 4}{\Sigma^{\prime}}^2-M_Q^2)
\{\xi^{\prime}_{\mu},\Sigma^{\prime }\}^2
- {i \over 2}{f^{(+)}}^{\prime}_{\mu
 \nu}{\xi^{\prime}}^{\mu}{\xi^{\prime}}^{\nu}\right.
\end{displaymath}
\be\label{A30}
\left.+ {1 \over 6} \left((d^{\prime}_{\mu}\xi^{\prime}_{\nu})^2
+ {1\over
2}(\xi^{\prime}_{\mu}{\xi^{\prime}}^{\mu}\xi^{\prime}_{\nu}{\xi^{\prime}}^{\nu}
 +
\xi^{\prime}_{\mu}\xi^{\prime}_{\nu}{\xi^{\prime}}^{\mu}
{\xi^{\prime}}^{\nu})\right)
-{1\over 6}d^{\prime}_{\mu}\Sigma^{\prime}
{d^{\prime}}^{\mu}\Sigma^{\prime}\right],
\ee

\noindent with

\be \label{A31}
(d^{\prime}_{\mu}\xi^{\prime}_{\nu})^2 =
-{1\over
4}[\xi^{\prime}_{\mu},\xi^{\prime}_{\nu}][{\xi^{\prime}}^{\mu},{\xi^{\prime}}^{
 \nu}]
+ i {f^{(+)}}^{\prime}_{\mu \nu} {\xi^{\prime}}^{\mu} {\xi^{\prime}}^{\nu}
+ {1\over 2}{f^{(-)}}^{\prime}_{\mu \nu}{f^{(-)}}^{{\prime}\mu \nu}
+ (d^{\prime}_{\mu} \xi^{\prime}_{\mu})^2 + \cdots,\ee

\be \label{A32}
\Gamma_{eff}^{(4)} = {N_c \over 16\pi^2} \Gamma(2,x)
tr\left[{1 \over 12}\xi^{\prime}_{\mu}\xi^{\prime}_{\nu}
{\xi^{\prime}}^{\mu}{\xi^{\prime}}^{\nu}
- {1 \over 6}\xi^{\prime}_{\mu}{\xi^{\prime}}^{\mu}
\xi^{\prime}_{\nu}{\xi^{\prime}}^{\nu}\right].\ee

When the shifts described in appendix B and diagonalization of quadratic terms
($\sigma \to \sigma + M_Q$ and $W^{(-)}_{\mu} \to
W^{(-)}_{\mu} + (1-g_A) \xi_{\mu}$)
are performed we are left with the effective action in terms of the
$0^-$, $1^-$, $1^+$ and $0^+$ fields.

\subsection{effective action for pseudoscalars.}

\begin{displaymath}
\Gamma_{eff}^{(2)} = {N_c \over 16\pi^2} \Gamma(0,x)<M_Q^2 g_A
\xi_{\mu} \xi^{\mu} + M_Q g_A^2\xi_{\mu} \xi^{\mu} \Sigma
+ M_Q i g_A \xi_{\mu} d^{\mu}\Delta
\end{displaymath}
\be\label{333}
+ {1 \over 48} (1-g_A^2)^2
[\xi_{\mu},\xi_{\nu}][\xi^{\mu},\xi^{\nu}] - {i \over 6} (1-g_A^2)
{f^{(+)}}_{\mu \nu} \xi^{\mu}\xi^{\nu}$$ $$-{1\over 12}{f^{(+)}}_{\mu
\nu}{f^{(+)}}^{\mu \nu} - {1\over 12} g_A^2 {f^{(-)}}_{\mu \nu}{f^{(-)}}^{\mu
\nu}>
\ee \
\begin{displaymath}
\Gamma_{eff}^{(3)} = {N_c \over 16\pi^2} \Gamma(1,x) <M_Q g_A^2\xi_{\mu}
\xi^{\mu} \Sigma - {i \over 2} g_A^2 {f^{(+)}}_{\mu \nu}
\xi^{\mu}\xi^{\nu} + {1 \over 8}g_A^2 (1-g_A^2)
[\xi_{\mu},\xi_{\nu}][\xi^{\mu},\xi^{\nu}]
\end{displaymath}
\begin{displaymath}
+ {1 \over 6}g_A^2 \left(-{1\over 4}[\xi_{\mu},\xi_{\nu}][\xi^{\mu},\xi^{\nu}]
+ {1\over 2}{f^{(-)}}_{\mu \nu}{f^{(-)}}^{\mu \nu}
+ i {f^{(+)}}_{\mu \nu}\xi^{\mu}\xi^{\nu}\right.
\end{displaymath}
\be\label{334}
\left.+ (d_{\mu} \xi^{\mu})^2 +
+ {1\over 2}g_A^2(\xi_{\mu}\xi^{\mu}\xi_{\nu}\xi^{\nu} +
\xi_{\mu}\xi_{\nu}\xi^{\mu}\xi^{\nu})\right)>
\ee
\
\be\label{335}
\Gamma_{eff}^{(4)} = {N_c \over 16\pi^2} \Gamma(2,x)
<{1 \over 12}g_A^4\xi_{\mu}\xi_{\nu}\xi^{\mu}\xi^{\nu}
- {1 \over 6}g_A^4\xi_{\mu}\xi^{\mu}\xi_{\nu}\xi^{\nu}>
\ee

\subsection{effective action for vectors.}

\begin{displaymath}
\Gamma_{eff}^{(2)} = {N_c \over 16\pi^2} \Gamma(0,x)
<-{1\over 12}{W^{(+)}}_{\mu\nu}{W^{(+)}}^{\mu\nu}
- {1\over 6}{W^{(+)}}_{\mu\nu}{f^{(+)}}^{\mu\nu}
\end{displaymath}
\be\label{336}
- {i\over 12}(1-g_A^2){W^{(+)}}_{\mu\nu}[\xi^{\mu},\xi^{\nu}]>
\ee

\be\label{337}
\Gamma_{eff}^{(3)} = {N_c \over 16\pi^2} \Gamma(1,x)
<-{i \over 6}g_A^2 {W^{(+)}}_{\mu\nu}[\xi^{\mu},\xi^{\nu}]>
\ee

\subsection{effective action for axial-vectors.}

\be\label{338}
\Gamma_{eff}^{(2)} = {N_c \over 16\pi^2} \Gamma(0,x)
< M_Q^2{W^{(-)}}_{\mu}{W^{(-)}}^{\mu} - {1 \over
 12}{W^{(-)}}_{\mu\nu}{W^{(-)}}^{\mu\nu}
- {1\over 6}g_A {W^{(-)}}_{\mu\nu}{f^{(-)}}^{\mu\nu}>
\ee
\be\label{339}
\Gamma_{eff}^{(3)} = {N_c \over 16\pi^2} \Gamma(1,x)
< {1 \over 12}{W^{(-)}}_{\mu\nu}{W^{(-)}}^{\mu\nu}
+ {1 \over 6}g_A {W^{(-)}}_{\mu\nu}{f^{(-)}}^{\mu\nu}>
\ee

\subsection{effective action for scalars.}

\be\label{341}
\Gamma_{eff}^{(1)} = {N_c \over 16\pi^2} \Gamma(-1,x)<2(\sigma+M_Q)^2 -
2\,M_Q^2
 + 2\,M_Q\Sigma + 2\,\sigma\Sigma>\ee

\be\label{342}
\Gamma_{eff}^{(2)} = {N_c \over 16\pi^2} \Gamma(0,x)
<d_{\mu}\sigma d^{\mu}\sigma + 2\,M_Q g_A^2\xi_{\mu} \xi^{\mu} \sigma -
4\,M_Q^2
 ( \sigma^2 + \sigma
\Sigma )>\ee

\be\label{343}
\Gamma_{eff}^{(3)} = {N_c \over 16\pi^2} \Gamma(1,x) < - 2\,M_Q g_A^2\xi_{\mu}
\xi^{\mu} \sigma - {2 \over 3} d_{\mu}\sigma d^{\mu}\sigma>\ee

\newpage

{\bf FIGURE CAPTIONS}

\vskip 2cm

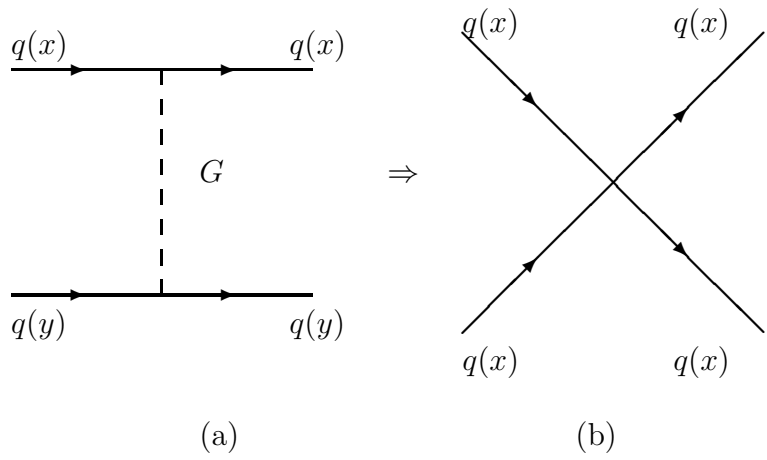
Figure 1:

(a) Conventional one-gluon exchange between two quark vertices in QCD.

(b) Local effective four-quark interaction which emerges
from (a) with the replacement in eq. (\protect{\ref{6a}}).

\vskip 2cm

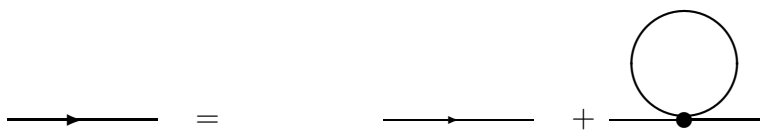
Figure 2:

Schwinger-Dyson equation for the quark propagator which leads
to the gap equation in eq. (\protect{\ref{69}}).

\vskip 2cm

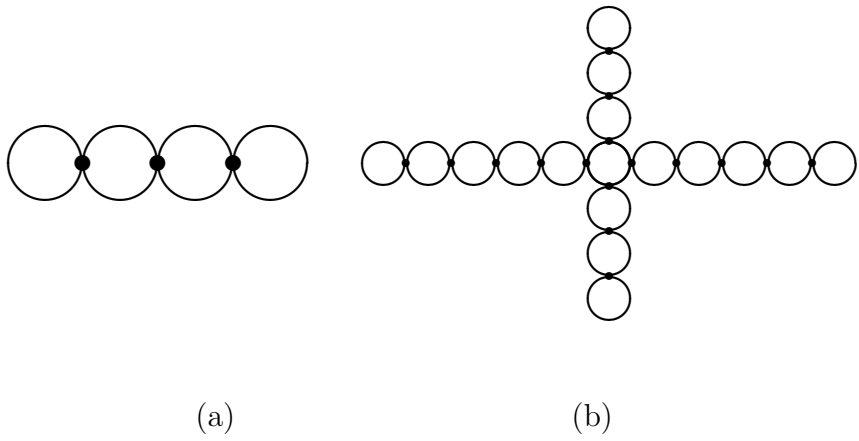
Figure 3:

In the leading large-$N_c$ approximation, the diagrams which
are summed are chains of fermion ``bubbles'' as in (a), as well as
trees of chains as in (b)

\vskip 2cm

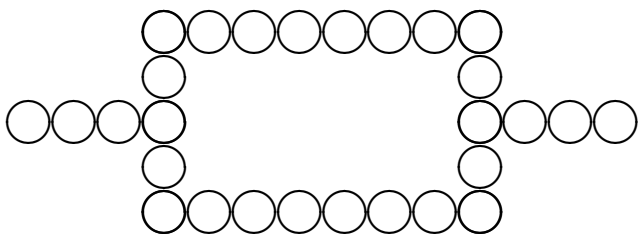
Figure 4:

An example of a loop of chains, which is next-to-leading
order in the $1/N_c$-expansion.

\newpage
\begin{figure}[htb]
\label{figure1}
\vspace*{2cm}
\unitlength 1cm
\begin{picture}(10,5)(-5,-2.5)
\thicklines
\put(-5,1.5){\vector(1,0){1}}
\put(-4,1.5){\vector(1,0){2}}
\put(-2,1.5){\line(1,0){1}}
\put(-5,-1.5){\vector(1,0){1}}
\put(-4,-1.5){\vector(1,0){2}}
\put(-2,-1.5){\line(1,0){1}}
\multiput(-3,1.5)(0,-0.4){8}{\line(0,-1){0.2}}
\put(-2.5,0){$G$}
\put(-5,-2){$q(y)$}
\put(-1.3,-2){$q(y)$}
\put(-5,1.7){$q(x)$}
\put(-1.3,1.7){$q(x)$}
\put(-2.5,-3.5){(a)}

\put(0,0){\Large $\Rightarrow$}

\put(1,-2){\vector(1,1){1}}
\put(2,-1){\vector(1,1){2}}
\put(4,1){\line(1,1){1}}
\put(1,2){\vector(1,-1){1}}
\put(2,1){\vector(1,-1){2}}
\put(4,-1){\line(1,-1){1}}
\put(1,2){$q(x)$}
\put(3.8,2){$q(x)$}
\put(1,-2.5){$q(x)$}
\put(3.8,-2.5){$q(x)$}
\put(2.5,-3.5){(b)}

\put(-0.5,-5){Fig. 1}

\end{picture}

\vskip 5cm

\end{figure}
\label{figure2}
\begin{figure}[htb]
\unitlength 1cm
\begin{picture}(10,3)(-5,-0.5)
\thicklines
\put(-5,0){\vector(1,0){1}}
\put(-4,0){\line(1,0){1}}
\put(-2.5,-0.1){=}
\put(4,0){\line(1,0){1}}
\put(4,0.75){\circle{1.5}}
\put(4,0){\circle*{0.2}}
\thinlines
\put(3,0){\line(1,0){1}}
\put(2.5,-0.1){+}
\put(0,0){\vector(1,0){1}}
\put(1,0){\line(1,0){1}}

\put(-0.5,-2){Fig. 2}

\end{picture}
\end{figure}

\newpage
\begin{figure}[htb]
\label{figure3}
\vspace*{2cm}
\unitlength 1cm
\begin{picture}(10,5)(-5,-2.5)
\thicklines
\multiput(-4.5,0)(1,0){4}{\circle{1}}
\multiput(-4,0)(1,0){3}{\circle*{0.2}}
\put(-2.5,-3.5){(a)}

\multiput(0,0)(0.6,0){11}{\circle{0.6}}
\multiput(0.3,0)(0.6,0){10}{\circle*{0.1}}
\multiput(3.0,-1.8)(0,0.6){7}{\circle{0.6}}
\multiput(3.0,-1.5)(0,0.6){6}{\circle*{0.1}}
\put(2.5,-3.5){(b)}

\put(-0.5,-5){Fig. 3}

\end{picture}
\end{figure}

\vskip 5cm

\begin{figure}[htb]
\label{figure4}
\unitlength 1cm
\begin{picture}(10,5)(-5,-2.5)
\thicklines
\multiput(-4,0)(0.6,0){4}{\circle{0.6}}
\multiput(-2.2,-1.2)(0,0.6){5}{\circle{0.6}}
\multiput(2.0,-1.2)(0,0.6){5}{\circle{0.6}}
\multiput(-2.2,-1.2)(0.6,0){8}{\circle{0.6}}
\multiput(-2.2,1.2)(0.6,0){8}{\circle{0.6}}
\multiput(2.0,0)(0.6,0){4}{\circle{0.6}}

\put(-0.5,-4){Fig. 4}

\end{picture}
\end{figure}

\end{document}